\documentclass[11pt,a4paper]{article}
\pdfoutput=1
\usepackage{jheppub}
\usepackage{float}
\usepackage{rotating}
\usepackage{array}
\usepackage[utf8]{inputenc}
\usepackage{amsmath}
\usepackage[normalem]{ulem}
\usepackage{slashed}
\usepackage{booktabs}
\usepackage[pdftex,table,dvipsnames]{xcolor}
\usepackage{units}
\usepackage{xfrac}
\usepackage{mathtools}
\usepackage{empheq}

\usepackage{multirow}

\newcommand{\be}{\begin{equation}}
\newcommand{\ee}{\end{equation}}
\newcommand{\bea}{\begin{eqnarray}}
\newcommand{\eea}{\end{eqnarray}}
\newcommand{\ds}{{\sf DarkSUSY}}

\preprint{DESY-19-140}

\title{Direct detection and complementary constraints for sub-GeV dark matter}

\author[a]{Kyrylo Bondarenko,}
\author[a]{Alexey Boyarsky,} 
\author[b]{Torsten Bringmann,}
\author[c]{Marco Hufnagel,}
\author[c]{Kai Schmidt-Hoberg}
\author[b]{and Anastasia Sokolenko}

\affiliation[a]{Intituut-Lorentz, Leiden University, Niels Bohrweg 2, 2333 CA, Leiden, The Netherlands}
\affiliation[b]{Department of Physics, University of Oslo, Box 1048, N-0371 Oslo, Norway}
\affiliation[c]{DESY, Notkestra\ss e 85, D-22607 Hamburg, Germany}

\emailAdd{bondarenko@lorentz.leidenuniv.nl}
\emailAdd{boyarsky@lorentz.leidenuniv.nl}
\emailAdd{torsten.bringmann@fys.uio.no}
\emailAdd{marco.hufnagel@desy.de}
\emailAdd{kai.schmidt-hoberg@desy.de}
\emailAdd{anastasia.sokolenko@fys.uio.no}

\abstract{
Traditional direct searches for dark matter, looking for nuclear recoils in deep underground detectors,
are challenged by an almost complete loss of sensitivity for light dark matter particles.
Consequently, there is a significant effort in the community to devise new methods and experiments to overcome these
difficulties, constantly pushing the limits of the lowest dark matter mass that can be probed this way.
From a model-building perspective, the scattering of sub-GeV dark matter on nucleons essentially 
must proceed via new light mediator particles, given that collider searches place extremely stringent 
bounds on contact-type interactions. Here we present an updated compilation of relevant limits for the 
case of a scalar mediator, including a new estimate of the near-future sensitivity of the NA62 experiment
as well as a detailed evaluation of the model-specific limits from Big Bang nucleosynthesis.
We also derive updated and more general limits on DM particles upscattered by cosmic rays, applicable to
arbitrary energy- and momentum dependences of the scattering cross section. 
Finally we stress that dark matter self-interactions, when evaluated beyond the common $s$-wave 
approximation, place stringent limits independently of the dark matter
production mechanism. These are, for the relevant parameter space, generically comparable to those that 
apply in the commonly studied freeze-out case.
We conclude that the combination of existing (or expected) constraints from accelerators and astrophysics, 
combined with cosmological requirements, puts robust limits on the maximally possible nuclear scattering rate. 
In most regions of parameter space these are at least 
competitive with the best projected limits from currently planned direct detection experiments.
}

\keywords{}

\begin{document}
\maketitle


\vspace*{-0.2cm}
\section{Introduction}

So far no unambiguous signal for new physics at the electroweak scale
has been identified at the Large Hadron Collider (LHC)~\cite{Aaboud:2017vwy,Sirunyan:2018vjp},
despite seemingly intriguing theoretical arguments that have been brought forward 
why the appearance of new physics should be expected at these energies 
(with low-scale supersymmetry being the most popular example, see e.g.~\cite{Craig:2013cxa}).
In consequence, while some natural islands remain~\cite{Dimopoulos:2014aua,Ross:2017kjc}, 
the experimental focus in the search for physics 
beyond the standard model of particle physics (SM) presently undergoes
a substantial broadening in scope, both concerning energy scales and theoretical 
frameworks for such searches~\cite{Beacham:2019nyx}. At the intensity frontier,
in particular, there is a plethora of both ongoing and planned activities that aim to explore new
physics in the sub-GeV range. Prominent examples for the latter include, but are not limited to, planned upgrades to current experiments such as NA62~\cite{NA62:2017rwk}  and NA64~\cite{Banerjee:2016tad},
the recently approved LHC add-on FASER~\cite{Feng:2017uoz,Ariga:2018pin,Ariga:2018uku,Ariga:2018zuc,Ariga:2019ufm}
as well as dedicated new experiments like LDMX~\cite{Akesson:2018vlm} and SHiP~\cite{Alekhin:2015byh,Anelli:2015pba,SHiP:2018yqc} that are planned to be run at the new Beam Dump Facility at CERN.
Finally there are proposals for LHC based intensity frontier experiments such CODEX-b~\cite{Gligorov:2017nwh} and MATHUSLA~\cite{Chou:2016lxi,Evans:2017lvd,Curtin:2018mvb,Lubatti:2019vkf}.

The existence of dark matter (DM) is one of the main arguments to expect physics beyond the SM. 
Also in this case theoretical considerations seem to point to the 
electroweak scale~\cite{Lee:1977ua}, independently of the arguments mentioned above, 
but direct searches for DM in the form of weakly interacting massive particles (WIMPs) have started
to place ever more stringent constraints on this possibility~\cite{Aprile:2018dbl,Ren:2018gyx}. Significant interest, both
from the experimental and theoretical perspective, has thus turned to the possibility of  
DM particle masses below the GeV -- TeV range. Conventional direct detection experiments 
are essentially insensitive to such light particles -- except for very large scattering cross sections, 
where cosmic rays can upscatter DM to relativistic energies~\cite{Bringmann:2018cvk} -- 
but new methods and concepts are being developed to overcome these difficulties~\cite{Hochberg:2015fth,Hochberg:2015pha,Acanfora:2019con,Dror:2019onn,Caputo:2019cyg,Dror:2019dib}. 

Both approaches may obviously be connected in terms of the underlying 
new physics, an insight which motivated a large body of phenomenological work studying possible complementary 
approaches to the DM puzzle (see, e.g., refs.~\cite{Batell:2009di,Essig:2009nc,Battaglieri:2017aum,Beacham:2019nyx,Lin:2019uvt},
and references therein).
In particular, the same new light messengers that are being probed at the intensity 
frontier could mediate interactions between the DM particles~\cite{Fayet:2006sp,Fayet:2007ua},
naturally leading to hidden sector freeze-out~\cite{Pospelov:2007mp,Batell:2009yf} as well as 
astrophysically relevant DM self-interactions~\cite{Feng:2009mn,Buckley:2009in,Tulin:2013teo}.
Recent discussions of complementary probes with a particular focus on light dark matter include 
refs.~\cite{Kouvaris:2014uoa, Krnjaic:2015mbs,Kainulainen:2015sva,Green:2017ybv,
Knapen:2017xzo,Evans:2017kti,Krnjaic:2017tio,Matsumoto:2018acr,Fradette:2018hhl}.
One of the main goals of this article is to further explore this connection.
%
The decisive link that allows to translate limits from searches for DM to those
for new particles that directly interact with the SM, and vice versa, is cosmology. 
It is worth stressing that, for a given model, stringent and robust cosmological bounds 
can typically be derived that are much less uncertain than general prejudice, or a model-independent 
assessment, would suggest. Throughout this work we therefore emphasise the need to 
consistently treat the non-trivial cosmological aspects appearing in scenarios with light mediators,
and base our limits on such a refined treatment.
In particular, we evaluate in detail the thermal evolution of the dark sector to 
compute the DM abundance and updated bounds from Big Bang Nucleosynthesis (BBN) 
-- but also demonstrate that DM self-interactions lead to stringent bounds that cannot
be evaded even if DM is not thermally produced via the common freeze-out mechanism.
We combine these new results with various updated accelerator constraints and projections,
and present them in a form directly usable by experimentalists probing the sub-GeV range.

This article is organised as follows. We start by reviewing the motivation to consider scenarios 
with light (scalar) mediators, in section \ref{sec:models}, and then introduce in more detail the case
of Higgs mixing that we will focus our analysis on. In section \ref{sec:dd} we present the 
current situation and near-future prospects of direct detection experiments, and  
generalise existing calculations for cosmic-ray accelerated DM to derive bounds on 
scattering cross sections involving light mediators. We then discuss particle physics 
constraints from various existing and planned experiments, in section \ref{sec:particle}, 
before investigating in detail the cosmological evolution of the dark sector in section \ref{sec:cosmo}.
In that section we also derive bounds from BBN and DM self-interactions that apply to
the specific scenario considered here, and mention further astrophysical bounds. In section \ref{sec:results} we
then combine the various constraints, and compare them to (projected) bounds from direct DM
detection experiments. Finally, in section \ref{sec:conc} we discuss our results 
and conclude.

\section{Models for light dark matter with portal couplings}
\label{sec:models}
Light dark sector particles are required to have small couplings to SM states in order to be allowed 
phenomenologically and therefore naturally correspond to fields that are singlets under the SM gauge interactions. 
They may then directly couple to the SM via the well known portal interactions~\cite{Batell:2009di}, 
i.e.~gauge-invariant and 
renormalisable operators involving SM and dark sector fields. If the DM particle $\chi$ is stable and fermionic, as assumed 
in this work, no direct renormalisable interaction is available and an additional particle $X$ mediating the interactions with 
the SM is required.\footnote{%
For scalar DM, on the other hand, there is a direct (Higgs) portal term~\cite{Silveira:1985rk}, 
constituting the most minimal DM model that is phenomenologically viable; for a recent status update see ref.~\cite{Athron:2017kgt}. 
Another portal term exists for a new heavy neutral lepton mixing 
with the SM model neutrinos; 
such a particle (often called `sterile neutrino') is not stable (decaying e.g.~into three SM 
neutrinos), but can be sufficiently long-lived to constitute DM~\cite{Boyarsky:2009ix}.
We do not consider these options here.
}
In recent years mediator searches at colliders together with complementary constraints from direct detection have therefore received a large amount of interest,
both for searches at the LHC~\cite{Frandsen:2012rk,Arcadi:2013qia,Buchmueller:2013dya,Garny:2014waa,Chala:2015ama,Jacques:2016dqz,Duerr:2016tmh}
and at low energy colliders~\cite{Schmidt-Hoberg:2013hba,Clarke:2013aya,Dolan:2014ska}. 
When comparing the sensitivities of collider searches with direct detection experiments it is important to take into 
account the large difference in energy scale between the centre of mass energy at colliders and the typical momentum transfer in nuclear recoils.
In particular the relative sensitivity of direct detection experiments is significantly increased for light mediators, implying that 
while scenarios with heavy mediators are strongly constrained by collider searches, those constraints are significantly weakened for light mediators.
Another appealing feature of light mediators, adding predictivity, 
is that DM can be produced within the standard thermal freeze-out paradigm\footnote{It has been noted that for heavy mediators 
DM overproduction can only be avoided in rather special corners of parameter space~\cite{Chala:2015ama,Jacques:2016dqz,Duerr:2016tmh}.}:
For sufficiently large couplings the dark sector will thermalise in the early universe and the DM relic abundance is 
set via annihilations into two mediators, 
$\chi \chi \rightarrow X X$, but also
via annihilations into SM fermions via an $s$-channel mediator,  $\chi \chi \rightarrow X \rightarrow \bar{f} f$,
if the dark sector is not fully decoupled at the time of freeze-out. 
Two particularly interesting and often studied options are vector mediators kinetically mixed with the SM 
hypercharge gauge boson or scalar mediators with Higgs mixing. 

\subsection{Vector mediators}

Let us start with a brief discussion of the vector mediator case. In the simplest scenario the field content consists of 
only a dark matter fermion charged under a dark $U(1)_X$ with kinetic mixing (see e.g.~\cite{Pospelov:2007mp, Essig:2009nc, Frandsen:2011cg}).
For light mediators the coupling structure will basically be that of a photon, so that $X$ predominantly decays to 
charged SM fermions such as electron positron pairs.
An important observation is that DM annihilations proceed via $s$-wave for both channels discussed above. If DM 
was ever in thermal contact with the SM (not necessarily through the kinetic mixing) such that the dark sector 
temperature is not much smaller than the photon temperature, there are strong constraints from Cosmic Microwave 
Background (CMB) observations, ruling out DM masses $m_\chi \lesssim 10$~GeV~\cite{Ade:2015xua}.
In fact these bounds extend to significantly higher DM masses for mediators parametrically lighter than DM due 
to the Sommerfeld enhancement of the annihilation cross section~\cite{Bringmann:2016din}.

There are a number of ways to evade this CMB limit, but they do involve some non-minimal component in the 
DM model. For instance DM may be asymmetric with only a sub-leading symmetric component 
such that residual annihilations during CMB are sufficiently suppressed~\cite{Baldes:2017gzu}. For consistency such a setup will 
however require the existence of an additional dark sector state to compensate the charge of the DM 
(reminiscent of electrons and protons). Another possibility would be to introduce a scalar whose vacuum 
expectation value (vev) generates 
small Majorana mass terms for the DM fermion, resulting in two dark matter states with slightly different 
mass, coupled off-diagonally to the vector boson (this is often referred to as inelastic DM)~\cite{Izaguirre:2015zva}. If the 
heavier state decays before the time of the CMB, s-wave annihilations $\chi_1 \chi_2 \rightarrow X \rightarrow \bar{f}f$ 
are no longer possible and constraints are evaded. A third possibility would be to couple the vector mediator to a light 
hidden sector state such that the decays of $X$ are invisible~\cite{Bringmann:2013vra}, in which case the CMB bounds can also 
be evaded. Finally, if the abundance is set via freeze-in~\cite{Bernal:2015ova} rather than freeze-out, the annihilation cross
section may be sufficiently small to be in accord with observations.

While all these options are viable and possess an interesting phenomenology, we wish to concentrate on a minimal 
setup in the current study. As discussed below, a model for light DM which still survives in its simplest form is that 
of a scalar mediator with Higgs mixing.

\subsection{Scalar mediators with Higgs mixing}
\label{sec:scalar}

In contrast to the case of a vector mediator, DM annihilations proceed via $p$-wave for a scalar mediator and 
the setup is correspondingly much less constrained by residual annihilations during CMB times.
If the dark sector was in thermal contact with the SM heat bath at even earlier times, however, dark sector masses are typically 
still required to be larger than $m_\chi \gtrsim$~10 MeV in order not to spoil the agreement between
predicted and observed primordial abundances of light nuclei (we will study the relevant limits in detail below). 

Let us consider a new real scalar $S$ that mixes with the SM Higgs and further couples to a new Dirac fermion $\chi$ that 
can play the role of the DM particle (see e.g.~ref.~\cite{Krnjaic:2015mbs} and references therein),
\begin{equation}
    \mathcal{L}_{S/\chi}
    = \frac{1}{2}\partial_\mu S \partial^\mu S + \bar{\chi}(i\slashed{\partial} - m_{\chi})\chi  - g_{\chi} S \bar{\chi} \chi -V(S,H).
    \label{eq:lagrangian-scalar}
\end{equation}
Here $m_{\chi}$ is the mass of the DM fermion, $H$ is the Higgs doublet of the SM and $V(S,H)$ is the scalar potential.
 The terms involving the singlet scalar can be written as 
 \begin{align}
 V(S,H) &=     \left(A_{hs} S + \lambda_{hs} S^2 \right) H^\dagger H + \mu_h^2 H^\dagger H + \lambda_h (H^\dagger H)^2  + V(S) \,.
 \end{align}
 with $V(S)= \xi_s S + 1/2 \mu_s^2 S^2  + 1/3 A_s S^3 + 1/4 \lambda_s S^4$.
 Without loss of generality the field $S$ can be shifted such that it does not obtain a vev, implying 
 $\xi_s = A_{hs} {v^2}/{2}$ (where the Higgs vev is given by $v=\left(\sqrt{2} G_F\right)^{-1/2}\simeq246.2$\,GeV). 
 After electroweak symmetry breaking the singlet $S$ mixes with the physical component of $H$ 
 such that the singlet $S$ naturally acquires a coupling to all SM fermions while the Higgs $h$ acquires a coupling to $\chi$, 
 \begin{equation}
 \mathcal{L} \supset - \sin \theta \frac{m_f}{v} S \bar{f}f - \sin \theta g_{\chi} h \bar{\chi} \chi 
 \label{eq:lagrangian_phi_chi}
 \end{equation}
 with mixing angle
 \begin{equation}
 \tan 2 \theta = \frac{2 A_{hs} v}{\mu_s^2-2 \lambda_h v^2}   \,.
 \end{equation}
The usual Higgs quartic coupling $\lambda_h$ is fixed in the SM via the observed Higgs mass
and we are interested in the parameter region $m_h \gg m_S$.
In our convention where $S$ does not acquire a vev the mixing angle is therefore approximately fixed by $A_{hs}$.
While the mixing angle clearly has a very large impact on most of the experimental observables, 
it does not fully specify the phenomenology of the scalar sector.
For example, the decay width of the SM-like Higgs boson into two light singlets is determined by the $S^2 H^\dagger H$ coupling,  
\begin{equation}
\Gamma_{SS} = \frac{\lambda_{hs}^2 v^2}{8\pi m_h}\sqrt{1-\frac{4 m_S^2}{m_h^2}} \,.
\label{eq:GammaSS}
\end{equation}
When we evaluate constraints e.g.\ from the Higgs signal strength we will assume that $\lambda_{hs} \simeq 0$ to be conservative.
Similarly we do not rely on $\lambda_{hs}$ for the thermalisation of the SM with the dark sector, yielding conservative limits from BBN.
We also assume the trilinear coupling $A_s$ to be small, so that the $3 \rightarrow 2$ annihilation rate of singlet scalars 
is negligible and no phase of `cannibalism' \cite{Farina:2016llk} occurs after freeze-out, again leading to conservative bounds.

For the calculation of DM-nucleus scattering rates we will also need the effective Yukawa coupling between a nucleon and the scalar mediator,
\begin{equation}
    g_{n,p} = \frac{m_{n,p}}{v} \sin\theta \left(\sum_{q=u,d,s} f_{q} + \frac{2}{9} f_G\right)\,. 
    \label{eq:effective_interaction_p}
\end{equation}
Here the constants $f_{q,G}$ correspond to the quark and gluon content of the nucleon.
It is well known that the couplings to protons and neutrons are very similar for Higgs exchange with $g_n \approx g_p \approx 1.16 \cdot 10^{-3} \sin\theta$, using state-of-the-art values for the $f_q$~\cite{Boveia:2016mrp}.

\section{Constraints from direct dark matter searches}
\label{sec:dd}

\subsection{Conventional light dark matter detection}

Direct detection experiments probe the elastic scattering 
cross section $\sigma_{\chi N}^\mathrm{SI}$ between DM particles $\chi$ and nuclei 
$N$ (since we only consider scalar mediators, we restrict our discussion to spin-independent scattering) at finite (spatial) momentum transfer 
\be
\label{eq:q2}
Q^2=2m_N T_N>0\,,
\ee
where $T_N$ is the nuclear recoil energy. 
For better comparison, however, these
results are typically reported in terms of the inferred cross section {\it per nucleon}, $\sigma_\mathrm{SI}$,
at {\it zero} momentum transfer. 
An assumption that is often adopted for the sake of this translation
is that of isospin-conserving couplings, which is almost perfectly satisfied for a scalar with Yukawa-like coupling structure.
This leads to the familiar coherent enhancement of
\be
\label{eq:coherence}
\sigma_{\chi N}^\mathrm{SI}(Q^2=0)=\sigma_\mathrm{SI} \times A^2 \frac{\mu_{\chi N}^2}{\mu_{\chi p}^2}\,.
\ee
Here $\mu_{\chi N}$ and $\mu_{\chi p}$ are the reduced masses of the DM/nucleus and DM/nucleon system,
respectively, and $A$ is the atomic mass number of the nucleus $N$.
Compared to this, the cross section at finite momentum transfer is suppressed by
a nuclear form factor $G_N$,
\be
\label{eq:GN}
\sigma_{\chi N}^\mathrm{SI}(Q^2)=\sigma_{\chi N}^\mathrm{SI}(Q^2=0)\times G_N^2(Q^2)\,.
\ee
This form factor is conventionally computed as the Fourier transform of the nuclear density profile,
i.e.~under the assumption that the scattering on the {\it nucleons} does not induce an additional momentum dependence.

For an interaction mediated by a scalar $S$ it is straightforward to calculate the non-relativistic scattering cross section
as~\cite{Krnjaic:2015mbs} 
\be
\label{sig0simp}
\sigma_{\chi N}^\mathrm{SI}(Q^2=0)=\frac{g_\chi^2 g_N^2 \mu_{\chi N}^2}{\pi m_S^4}\,,
\ee
where $g_N$ denotes the coupling between $S$ and the nucleus, i.e.~$g_N=A g_p$ if isospin is conserved. 
In the case of Higgs mixing, using eqs.~(\ref{eq:effective_interaction_p}), (\ref{eq:coherence}) and (\ref{sig0simp}), this translates 
to the DM-proton scattering cross section  
\be
\label{eq:sigsifull}
   \sigma_\mathrm{SI} = 1.7\cdot 10^{-34}\,\text{cm}^2\times g_{\chi}^2 \sin^2\theta \left(\frac{m_{S}}{\text{GeV}}\right)^{-4} \left(\frac{m_{\chi}}{\text{GeV}}\right)^2
    \left(1 + \frac{m_{\chi}}{m_p}\right)^{-2}.
 \ee
For a heavy mediator, this expression can directly be compared to standard limits on  $\sigma_\mathrm{SI}$
because scattering on nucleons is essentially momentum-independent.
If the mediator is light compared to the typical momentum transfer, however, 
the cross section probed in the detector is smaller than expected from eq.~(\ref{eq:GN}) and 
limits have hence to be re-evaluated taking into account all the relevant experimental information. 
An approximate -- but still  reasonably accurate -- way of taking into account the 
momentum suppression consists in simply rescaling
(see, e.g., ref.~\cite{Kaplinghat:2013yxa})
\be
\label{simp_resc}
 \sigma_{\chi N}(Q^2=0)\simeq  \tilde \sigma_{\chi N}(Q^2=0)\times \frac{(m_S^2+Q^2_\mathrm{ref})^2}{m_S^4}\,,
\ee
where $\tilde \sigma_{\chi N}$ is the limit reported under the assumption of a constant scattering cross section 
-- in terms of $\sigma_\mathrm{SI}$ as given in 
eq.~(\ref{eq:coherence}) -- and $Q^2_\mathrm{ref}$ is an experiment-specific reference momentum transfer
(see also ref.~\cite{Kahlhoefer:2017ddj} for a discussion of how to explore light mediators with direct detection 
experiments).

\begin{figure}[t]
\begin{center}
\begin{minipage}{0.01\textwidth}
\mbox{ }~
\end{minipage}
\begin{minipage}{0.23\textwidth}
\footnotesize
   \vspace{0.15cm}
   \begin{tabular}{|l|c|}
        \hline
         & \small  $\left(Q_\mathrm{ref}^2\right)^{1/2}$ [MeV] \\ \hline\hline
          CRESST-III~\cite{Abdelhameed:2019hmk}  & $3.2$ \\ \hline 
        DarkSide-50~\cite{Agnes:2018ves}  & $6.7$ \\ \hline
        PandaX-II~\cite{Ren:2018gyx}  & $26$ \\ \hline
        Xenon 1T~\cite{Aprile:2018dbl} & $35$ 
        \\\hline\hline
        DARWIN~\cite{Aalbers:2016jon} & $40$ \\ \hline 
         NEWS-G~\cite{Battaglieri:2017aum,Arnaud:2017bjh} & $1.5$ \\ \hline 
        SuperCDMS~\cite{Agnese:2016cpb} & $2.3$\\ \hline 
        LUX-ZEPLIN~\cite{Akerib:2018lyp} & $16$\\ \hline 
    \end{tabular}
\end{minipage}
\begin{minipage}{0.2\textwidth}
\mbox{ }\hspace*{\textwidth}
\end{minipage}
\begin{minipage}{0.53\textwidth}
\vspace*{0.5cm} \mbox{ }\\
\includegraphics[width=\textwidth]{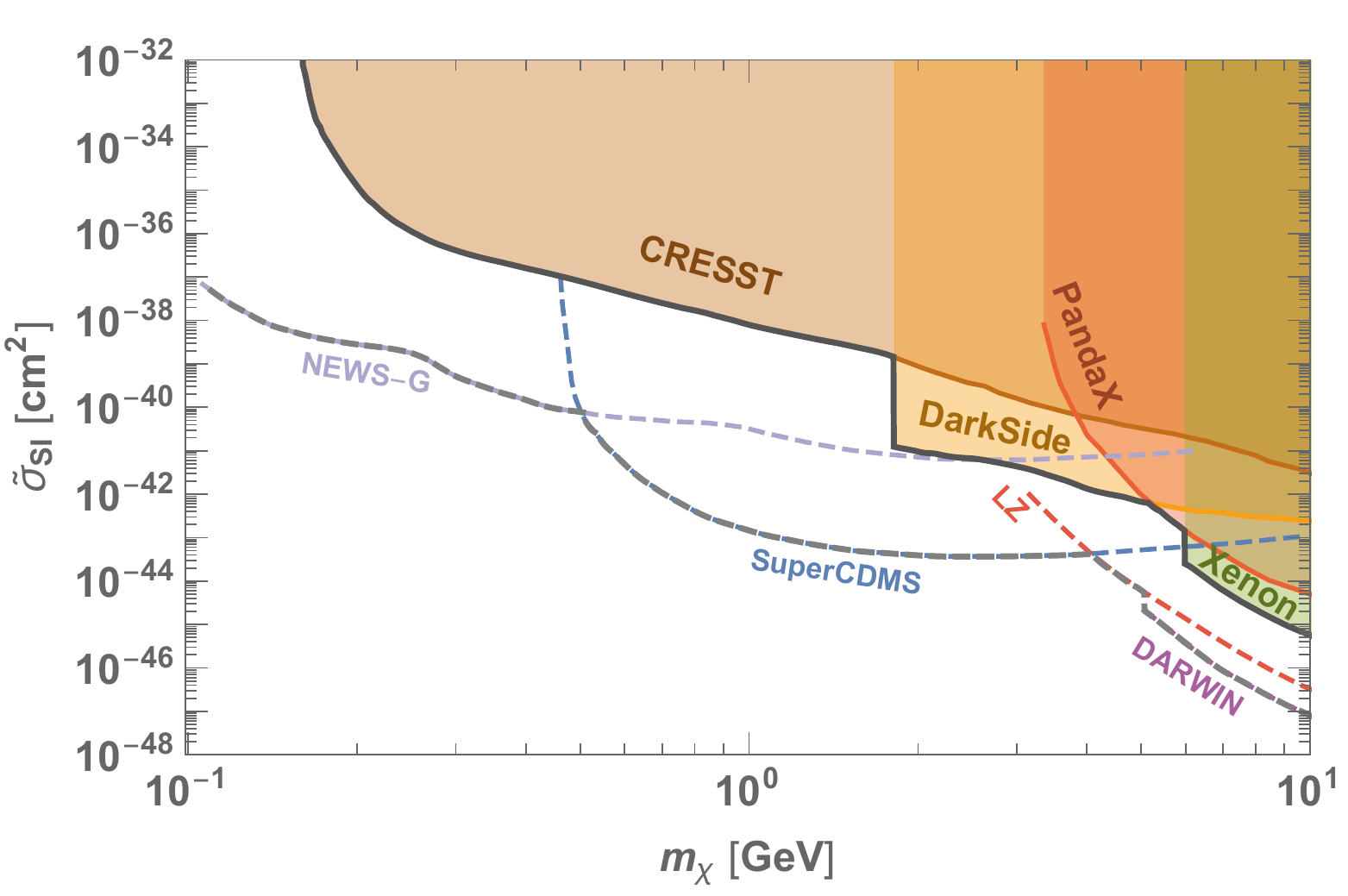}
\end{minipage}
\vspace*{-0.5cm}
\end{center}
\caption{{\it Left panel}. Current (upper part) and future (lower part) direct detection experiments, along with a 
reference scale for the momentum transfer. 
{\it Right panel}. Current limits (solid lines) and projected sensitivities (dashed lines) to the nucleon cross section 
{\it assuming momentum-independent scattering}. Limits for the zero momentum
limit are thus obtained by re-scaling these reported results as in eq.~(\ref{simp_resc}).
\label{Fig:DDexp}}
\end{figure}

In figure~\ref{Fig:DDexp} we summarise the most stringent (projected) direct detection constraints at low DM 
masses, along with the value of $Q^2_\mathrm{ref}$ that we use for the corresponding experiment. 
The latter was either estimated by using eq.~(\ref{eq:q2}) for the minimal recoil energy adopted in the respective 
analysis, 
or by directly fitting to data provided by the experiment (for PandaX-II~\cite{Ren:2018gyx}).
We note that carefully modelling inelastic scattering processes, resulting in the emission of a 
photon or an atomic electron, in principle allows to improve sensitivities
in the few 100\,MeV range~\cite{Kouvaris:2016afs,Ibe:2017yqa,Dolan:2017xbu}. 
There is also a number of proposed direct detection experiments, and ideas, that would probe
even smaller cross sections in the mass range shown in figure~\ref{Fig:DDexp}, but the status of those is 
presently less certain (for a recent compilation, see ref.~\cite{Knapen:2017xzo,Battaglieri:2017aum}).

\subsection{Cosmic ray-accelerated dark matter}

The right panel of figure~\ref{Fig:DDexp} clearly illustrates the exponential loss of sensitivity of conventional direct 
detection experiments  to sub-GeV DM, reflecting the fact that non-relativistic DM particles 
with such small masses do not carry enough momentum to allow for nuclear recoils above the 
experimental threshold.
As recently pointed out, however, there is a small yet inevitable component of {\it relativistic} DM
that alleviates this limitation~\cite{Bringmann:2018cvk}:\footnote{%
A subdominant population of DM particles with velocities exceeding the galactic escape velocity has also 
been considered in Refs.~\cite{Agashe:2014yua, Kouvaris:2015nsa,An:2017ojc,Emken:2017hnp,Cappiello:2018hsu,Ema:2018bih,Alvey:2019zaa,Cappiello:2019qsw,Dent:2019krz}.
} 
if DM can elastically scatter with nuclei, then
also the well-established population of high-energy cosmic rays will scatter on DM, thus
accelerating them from essentially at rest (in the galactic frame) to GeV energies and beyond -- in 
principle for arbitrarily small DM masses.

In order to handle scattering via light mediators we extend the formalism developed in 
ref.~\cite{Bringmann:2018cvk} to allow for arbitrary relativistic scattering amplitudes (rather than only 
a constant $\sigma_{\chi N}$ as assumed there). As the derivation
follows the same steps as in ref.~\cite{Bringmann:2018cvk}, we only briefly state our results here and 
refer to that reference for further details (see also ref.~\cite{Dent:2019krz}). The flux of cosmic-ray 
accelerated DM (CRDM) before a potential 
attenuation in the Earth or the atmosphere is given by
\be
\label{dphidtdm}
{\frac{d\Phi_\chi}{dT_\chi} =  D_\mathrm{eff} \frac{\rho_\chi^\mathrm{local}}{m_\chi}  
\int_{T_{CR}^\mathrm{min}}^\infty d T_{CR}\, \frac{d \sigma_{\chi N} }{dT_\chi} \frac{{d\Phi^\mathrm{LIS}_{CR}}}{dT_{CR}} }\,.
\ee
Here, $\rho_\chi^\mathrm{local}$ and $\Phi^\mathrm{LIS}_{CR}$ are the local interstellar DM density and
the cosmic-ray flux, respectively, and ${T_{CR}^\mathrm{min}}$ is the minimal 
kinetic cosmic-ray energy needed to accelerate DM to kinetic energy $T_\chi$; we take into account elastic scattering
of cosmic-ray nuclei $N=\{p,\,^4\mathrm{He}\}$ with DM, including in each case the same 
dipole form factor suppression as in ref.~\cite{Bringmann:2018cvk}.\footnote{%
Note that this is a conservative estimate, neglecting inelastic DM-CR interactions, which will become 
relevant at sufficiently large values of the momentum transfer. 
We leave a detailed study of these effects for future work.}
$D_\mathrm{eff}\sim8$\,kpc, finally, is an effective distance out to which we assume that the source density of CRDM
is roughly the same as it is locally (which, for a standard DM distribution, corresponds to a sphere of about 10\,kpc diameter).
The scattering rate of relativistic CRDM particles in underground detectors is then determined as
\be
 {\frac{d\Gamma_N}{d T_{N}}= 
 \int_{T_\chi(T_\chi^{z, \mathrm{min}})}^\infty \!\!dT_\chi\ 
 \frac{d \sigma_{\chi N}}{dT_N} \frac{d\Phi_\chi}{dT_\chi}} \,,
\ee
where the scattering cross section ${d \sigma_{\chi N}}/{dT_N} $ must be evaluated for the actual DM energy
$T_\chi^z$ at the detector's depth $z$ (which is lower than the initial DM energy $T_\chi$ due to soil 
absorption~\cite{Starkman:1990nj,Mack:2007xj,Hooper:2018bfw,Emken:2018run}), 
and $T_\chi(T_\chi^{z, \mathrm{min}})$ denotes the minimal initial CRDM energy that is needed to induce
a nuclear recoil of energy $T_N$ (again taking into account a potential attenuation of the flux due to the propagation 
of DM through the Earth and atmosphere). 
In order to relate $T_\chi^z$ to the initial DM energy $T_\chi=T_\chi^{z=0}$, we numerically solve 
the energy loss equation
\be
\frac{dT_\chi^z}{dz}=-\sum_N n_N\int_0^{T_N^\mathrm{max}}\!\!\!dT_N\,\frac{d \sigma_{\chi N}}{dT_N} T_N\,,
\ee
where $T_N^\mathrm{max}$ denotes the maximal recoil energy $T_N$ of nucleus $N$, for a given DM
energy $T_\chi^z$, and we sum over the 11 most abundant elements in Earth's crust.

It is worth stressing that the momentum transfer in a direct detection experiment is given by eq.~(\ref{eq:q2}) also in 
the relativistic case. In particular, the form factor in the nuclear scattering cross section does not depend on the energy 
of the incoming DM particles, only on the relatively small range of $Q^2$ that falls inside the experimental target region. 
This makes it straightforward to translate direct detection limits reported in the literature for {\it heavy} DM, 
assuming the standard DM halo profile and velocity distribution,
to a maximal count rate in the analysis window of recoil energies and in turn
to limits resulting from the CRDM component discussed here~\cite{Bringmann:2018cvk}.
The updated routines for the computation of the resulting CRDM flux and underground scattering rates have been 
implemented in \ds~\cite{Bringmann:2018lay}, which we also use to 
calculate the resulting limits from a corresponding re-interpretation of Xenon-1T~\cite{Aprile:2018dbl} results.

In order to do so, we still need the full relativistic scattering cross section of DM with nuclei, mediated by
a scalar particle. 
For fermionic nuclei we find
\be
\label{diffsig_full}
\frac{d\sigma_{\chi N}}{d T_N}=\frac{\sigma_{\chi N}^\mathrm{SI, NR}}{16 \mu_{\chi N}^2sT_N^\mathrm{max}}
\frac{m_S^4}{(Q^2+m_S^2)^2}
(Q^2+4m_N^2)(Q^2+4m_\chi^2)\times G_N^2(Q^2)\,,
\ee
where $\sigma_{\chi N}^\mathrm{SI, NR}$ is the scattering cross section in the highly non-relativistic 
limit, as stated in eq.~(\ref{sig0simp}), $s=E_\mathrm{CM}^2$ and $G_N(Q^2)$ is the conventional nuclear form factor.
While the non-relativistic result is of course recovered for $Q^2\to0$ and $s\to(m_\chi+m_N)^2$,
this cross section is actually {\it enhanced} for $Q^2\gtrsim m_\chi^2$ when
compared to the standard estimate given in eq.~(\ref{simp_resc}). This is particularly relevant 
both for very light DM ($m_\chi^2\lesssim Q^2_\mathrm{ref}$) and
the {\it production} of the CRDM component stated in eq.~(\ref{dphidtdm}), for which the momentum 
transfer is typically much larger than expected in underground experiments.

\begin{figure}[t]
\begin{center}
\hspace*{-0.3cm}
\includegraphics[width=0.49\textwidth]{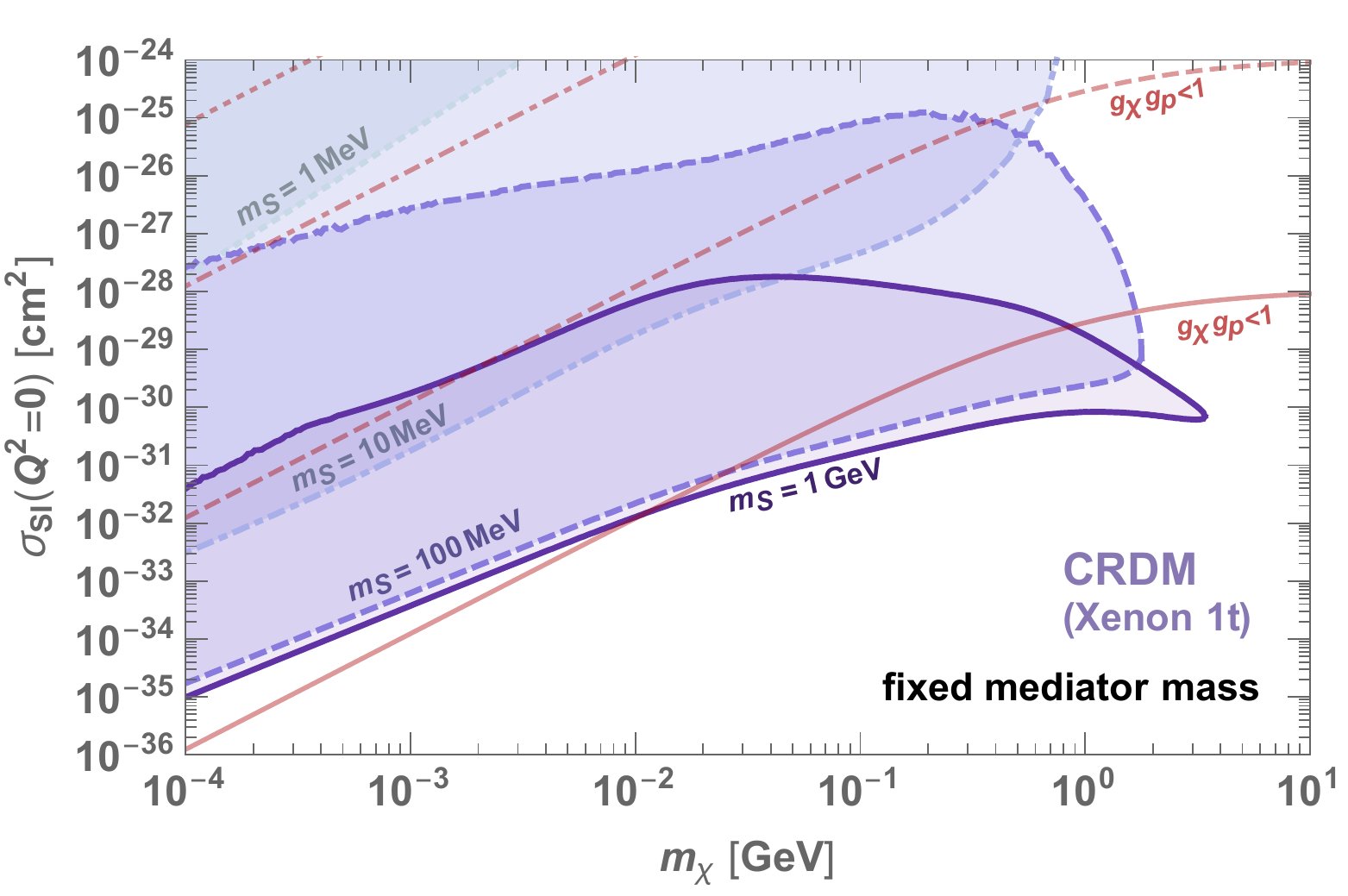}
\hspace*{0.2cm}
\includegraphics[width=0.49\textwidth]{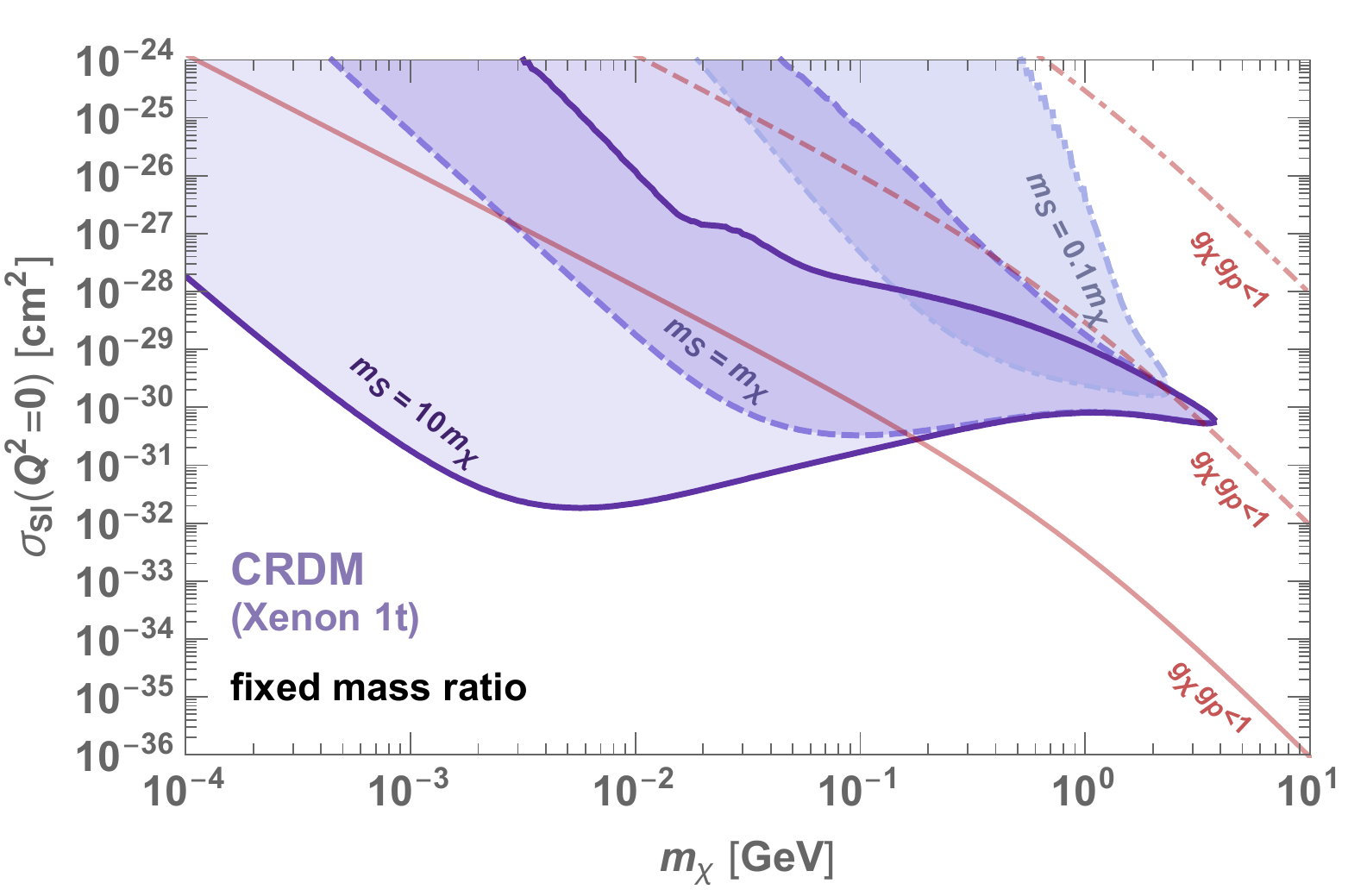}
\end{center}
\caption{{\it Left panel}. Direct detection constraints on dark matter accelerated by cosmic rays for fixed
mediator masses. Cross sections below the lower boundaries lead to recoil rates too small to be detectable,
while cross sections above the upper confining boundaries prevent the dark matter particles to
reach the detector, due to efficient scattering in the overburden. 
As a rough indication of how large cross sections are in principle possible, we also 
show in each case the parameter range where the couplings are well inside the perturbative regime 
(for a more detailed treatment, see ref.~\cite{Digman:2019wdm}).
{\it Right panel}. Same, for fixed mediator to DM mass ratios.}
\label{fig:CRDM_constraints}
\end{figure}

In figure~\ref{fig:CRDM_constraints} we show the resulting limits from Xenon-1T on light DM.
An important feature of a constant scattering cross section is that these constraints (almost)
flatten for very small DM masses~\cite{Bringmann:2018cvk}. Compared to that,
as expected from the above discussion (see also ref.~\cite{Dent:2019krz}), we observe a significant strengthening 
of our constraints at fixed mediator masses. 
However the figure also clearly demonstrates that for light mediator masses the production of the CRDM component
becomes suppressed by the mediator momentum; when considering only mediators that are lighter than the DM particle, 
in particular, the resulting constraints become less and less stringent.
Also the behaviour of the {\it maximal} cross section (due to soil absorption) 
is rather instructive, as it falls into two clearly distinguishable regimes: {\it i)} for heavy (GeV-scale and above)
mediators the upper boundary essentially follows that of the constant cross section case~\cite{Bringmann:2018cvk},
roughly rescaled by an additional $m_\chi^2$ dependence (for small $m_\chi$) with the same origin
as discussed above for the lower boundary; {\it ii)} for lighter mediator masses, the momentum suppression
starts to become relevant, strongly favouring scattering events in the overburden with small momentum 
transfers -- which in turn leads to a significantly increased penetration depth, and hence weaker constraints.

Let us, finally, stress that the limits presented in figure~\ref{fig:CRDM_constraints} in principle apply to any 
model with scalar mediators, i.e.~they are not restricted to the specific structure of the DM-nucleon
coupling given in eq.~(\ref{eq:effective_interaction_p}).

\section{Constraints from particle physics experiments}
\label{sec:particle}

Let us now turn to constraints on the scalar portal model from particle physics experiments. In the following we 
concentrate mostly on the case $m_S \lesssim m_\chi$
 so that the annihilation channel $\bar{\chi} \chi \rightarrow SS$ is kinematically allowed in the early universe.
The reason is that for $m_S \gtrsim m_\chi$ only direct annihilations into SM states via an $s$-channel 
scalar singlet are allowed, $\bar{\chi} \chi \rightarrow S \rightarrow \text{SM}$ (see section \ref{sec:evolution}
for a more detailed discussion). The corresponding 
annihilation rate, however, is typically constrained to be too small to allow for the observed DM relic abundance 
(see e.g.~\cite{Krnjaic:2015mbs}), making this case less appealing. Note that $m_S \lesssim m_\chi$ 
naturally implies that the singlet scalar $S$ can only decay to SM states that can potentially be 
observed in detectors (`visible decays'). Depending on the mixing angle $\theta$, however, the lifetime 
of $S$ can be so long that the decay happens only outside of the detector and the signature is therefore 
identical to an invisibly decaying scalar.
 While we mainly concentrate on this case, we will also briefly comment on the case $m_S \gtrsim m_\chi$.
 
An important property of the inherited Yukawa-like coupling structure is that the production of $S$ 
may well proceed via one of the larger Yukawa couplings, while its decay is typically controlled by 
smaller couplings because only light states are kinematically accessible. In particular, flavour 
changing transitions induced at the loop level are typically very relevant (see, e.g., \cite{Bird:2006jd}) 
and lead to the production via rare meson decays such 
as $B \rightarrow K S$ and $K \rightarrow \pi S$, which are strongly constrained by a variety of  
experiments~\cite{Bird:2004ts,Batell:2009jf,Alekhin:2015byh,Boiarska:2019jym}. Constraints on light scalars 
as well as projected sensitivities have been evaluated frequently in the literature, with a recent compendium 
of limits shown e.g.~in ref.~\cite{Winkler:2018qyg}
(see also ref.~\cite{Bezrukov:2009yw}, pointing out that such a light scalar could even drive cosmological inflation).
In addition invisible decays of the SM Higgs into DM, $h \rightarrow \bar{\chi}\chi$ can give relevant 
constraints on the same product of couplings, $g_{\chi}\cdot \sin \theta$, that is relevant for direct detection. 
In the following we briefly summarise the limits that we use in our analysis.

\subsection{Invisible Higgs decay and signal strength}
\label{sec:higgsinv}

Data on Higgs bosons created at the LHC in principle constrain the SM Higgs mixing angle $\theta$ 
in two ways. First, invisible Higgs decays are constrained as $\text{BR}(h \to \text{inv.}) < 0.19$ 
(95\% C.L.)~\cite{Sirunyan:2018owy}.
Second, the observed Higgs signal strength
 \begin{equation}
     \mu\equiv \frac{N_h^{\text{exp}}}{N_h^{\text{SM}}} = \frac{[\sigma_h\, \text{BR}(h\to \text{SM})]_{\text{exp}}}{[\sigma_h\,\text{BR} (h\to \text{SM})]_{\text{SM}}},
 \end{equation}
 where $\sigma_h$ is the Higgs boson production cross section and $N_h^{\text{exp},\text{SM}}$ is the number 
 of observed and expected Higgs bosons, respectively,
is constrained to be $\mu > 0.89$ (95\% C.L.)~\cite{AtlasCMSHiggsSignalStrength}. 
 In our model the latter constraint is more stringent because the Higgs boson production cross section 
 can only be reduced compared to the SM case, thus implying $\text{BR}(h \to \text{inv.}) < 0.11$.

Specifically there are three effects that lead to a reduction of the signal strength:
\begin{enumerate}
 \item Reduction of production cross section and decay widths for $h$, due to mixing.
 \item An invisible branching fraction, $h \rightarrow \bar{\chi} \chi$.
 \item Decays into two scalars, which depletes the branching ratio in the remaining channels. 
\end{enumerate}
In our case
the ratio of the production cross sections is simply given by
\begin{equation}
    \sigma_h^{\text{model}}/\sigma_h^{\text{SM}} = \cos^2\theta\,,
\end{equation}
and for the branching ratios we have
\begin{equation}
    \frac{\text{BR}^{\text{model}}(h\to \text{SM})}{\text{BR}^{\text{SM}}(h\to \text{SM})} = \frac{\cos^2 \theta \, \Gamma_0}{\cos^2 \theta \, \Gamma_0 + \Gamma_{SS}  + \Gamma_\text{inv}}\, .
\end{equation}
Here $\Gamma_0 \approx 4.1$~MeV is the total SM Higgs width (without mixing), 
\begin{equation}
\Gamma_\text{inv} = \frac{g_\chi^2 \, m_h \, \sin^2 \theta}{8 \pi} \left(1 - \frac{4 \, m_\chi^2}{m_h^2}\right)^{3/2}
\end{equation}
 is the partial decay width for invisible decays
 and $\Gamma_{SS}$ is the Higgs boson decay width into two scalars 
 (see eq.~\eqref{eq:GammaSS} and related discussion). 
Here we conservatively assume $\lambda_{hs}$ to be negligibly small, and hence set $\Gamma_{SS}\approx0$.
Taken together, the limit resulting from the predicted Higgs signal strength is thus given by
\begin{equation}
\label{eq:higgslimit}
 \mu_{\text{model}} = \cos^2 \theta \times \text{BR}^{\text{model}}(h \rightarrow \text{SM}) = \frac{\cos^4 \theta \, \Gamma_0}{\cos^2 \theta \, \Gamma_0  + \Gamma_\text{inv}}>0.89\,,
\end{equation}
which for $m_\chi \ll m_h$ implies
\begin{equation}
\label{eq:higgspresent}
\sin^2 \theta \, g_\chi^2 \lesssim 1.0 \cdot 10^{-4} \;.
\end{equation}
This limit will soon be improved with data from the 13~TeV run (see e.g.~\cite{Robens:2019kga}). 
For the high luminosity phase of the LHC
the direct bound on the invisible branching ratio will become more constraining and we use 
\begin{equation}
\label{eq:higgsfuture}
 \text{BR}_\text{inv}\simeq \frac{\Gamma_\text{inv} }{\Gamma_\text{inv} +\Gamma_0} <0.025
\end{equation}
as an estimate of the future bound~\cite{Cepeda:2019klc}, thus strengthening the bound in eq.~(\ref{eq:higgspresent})
by a factor of about 0.21.\footnote{The ILC could improve on this limit significantly, but we 
do not include the corresponding sensitivity 
as the status of the ILC is far from clear at this point.}

\newpage
\vspace*{-1.5cm}
\subsection{Limits on $\sin\theta$ from beam dumps and colliders}
\label{sec:S_constraints}

As already mentioned, singlet scalars $S$ can be efficiently produced via the decay of heavy mesons
which in turn are copiously  produced at the collision energies available at  SPS and LHC 
based intensity frontier experiments, see e.g.~Ref.~\cite{Bird:2006jd}.\footnote{%
If the value of the quartic coupling $\lambda_{hs}$ is sizeable, production via Higgs boson decays 
may  become dominant for  LHC based experiments and for $m_S>m_B$~\cite{Boiarska:2019vid}. 
We do not discuss this case here.
}
For scalars too heavy to be produced in $B$ meson decays this production mechanism is not available and 
the most constraining limit comes from LEP~\cite{Winkler:2018qyg}. 

For scalar masses kinematically accessible to  $B \rightarrow K S$  but too large to allow for $K \rightarrow \pi S$, 
the strongest constraints are typically obtained from experiments where both the scalars can be created and 
their decay products can be detected. For experiments of this type the number of detected events thus scales 
with $\sin^4\theta$ (at the lower bound of sensitivity). 
For our analysis we use the results from LHCb~\cite{Aaij:2016qsm} as well as a reinterpretation~\cite{Winkler:2018qyg} 
of the past beam-dump experiment CHARM~\cite{Bergsma:1985qz}.
Values of $\sin\theta$ just below the current sensitivity will be tested by LHCb in the high luminosity phase and we estimate the 
corresponding sensitivity (see appendix~\ref{app:lhcb} for details).
For smaller values of $\sin\theta$ 
the SHiP experiment~\cite{Alekhin:2015byh,SHiP:2018yqc,Bondarenko:2019yob} 
and MATHUSLA~\cite{Chou:2016lxi,Curtin:2018mvb} have the best sensitivities 
$m_S \lesssim 5$~GeV~\cite{Beacham:2019nyx}. 
Both experiments aim at working in the background-free regime (see Refs.~\cite{Anelli:2015pba,SHiP:2018yqc} for 
detailed simulations for SHiP and Refs.~\cite{Curtin:2017izq,Evans:2017lvd,Alpigiani:2018fgd,Curtin:2018mvb} for MATHUSLA). 
In reality, it is very difficult however to completely exclude the possibility of residual background events to take 
place in the detector. In case of SHiP such events can be distinguished from the signal events by making use of 
the spectrometer, mass reconstruction and particle identification. These options are not available in the case of 
MATHUSLA, implying that it is not straight-forward to compare its reported formal sensitivity (based on 2.3 events in the detector) 
to the one from SHiP. For this work we will therefore concentrate on the expected bounds from SHiP.

\begin{figure}[t]
\begin{center}
\vspace*{-0.5cm}
\includegraphics[width=0.6\textwidth]{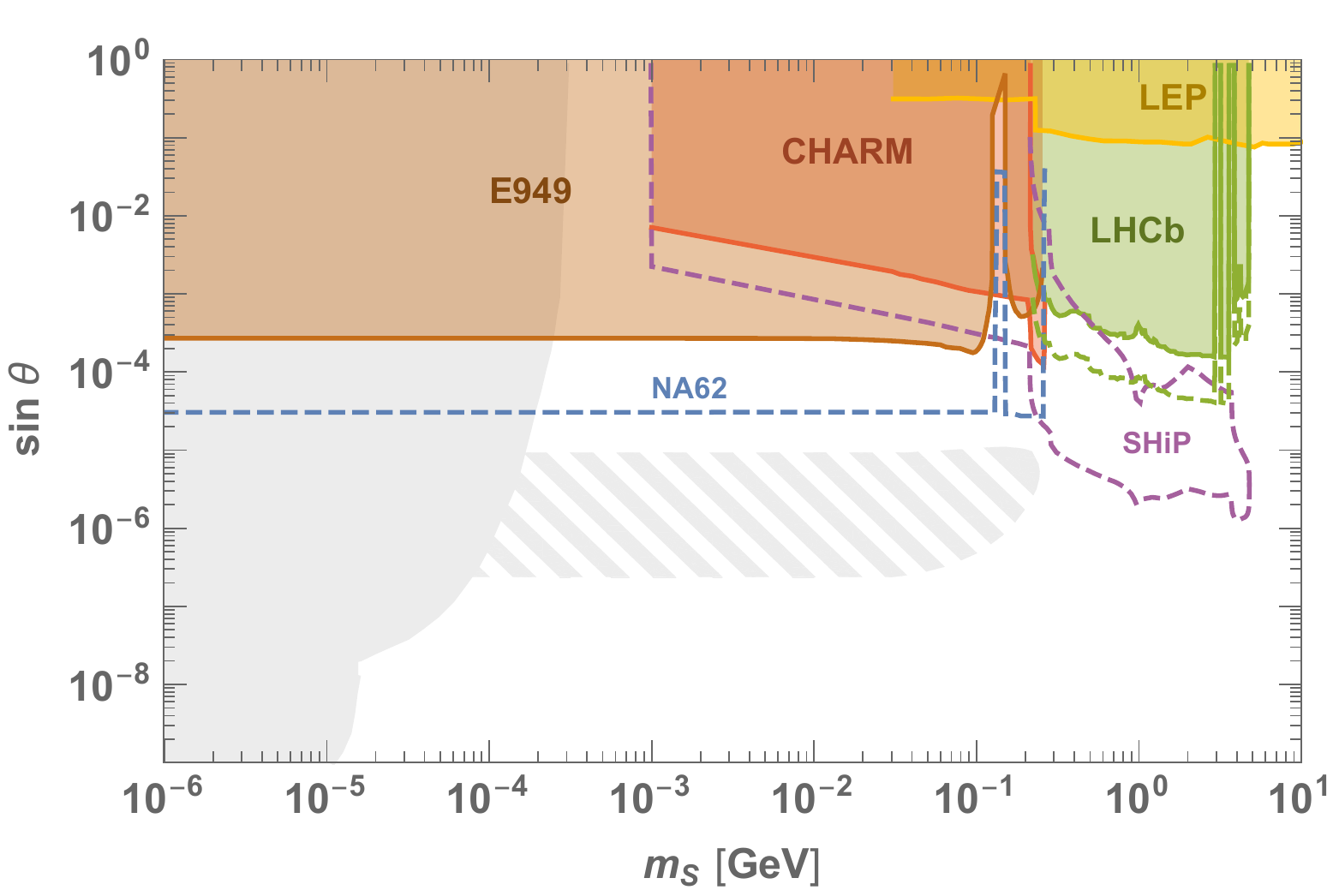}
\vspace*{-0.5cm}
\end{center}
\label{fig:astro-accel}
\caption{Current limits (solid lines) and projected sensitivities (dashed lines) 
from accelerator and beam dump searches for new light scalars $S$ decaying visibly into standard
model particles according to eq.~(\ref{eq:lagrangian_phi_chi}). See text for more details and references. 
For comparison, we also indicate the astrophysical constraints discussed in section \ref{sec:astro} (grey area).
}
\label{fig:particle_constraints}
\end{figure}

For smaller scalar masses, $m_S<m_K-m_{\pi}\approx 350\,$MeV, experiments that search for rare kaon decays are typically more 
sensitive. The reason is that, unlike for particles heavier than kaons, one can perform
precision measurements of the final state pion energy, on an event by event basis. 
The number of confirmed signal events thus no longer depends on the detection of the scalar decay products
and therefore scales as $\sin^2 \theta$, i.e.~is much less suppressed than in the case of heavier scalars. 
Both experiments E949~\cite{Artamonov:2009sz} and NA62~\cite{NA62:2017rwk} 
search for rare $K^+\to \pi^+ + \text{MET}$ decays and give bounds on the scalar mixing through the process 
$K^+\to \pi^+ S$. As this is independent of the decay of $S$, scalars with arbitrarily small masses can be constrained.
In addition to the limit from E949 we estimate the sensitivity of NA62 during LHC Run 3 (see 
appendix~\ref{app:NA62} for details); we note that the resulting sensitivity is very similar to the 
sensitivity of the proposed KLEVER experiment shown in ref.~\cite{Beacham:2019nyx}.
In figure~\ref{fig:particle_constraints} we show all current limits (full lines) as well as future sensitivities 
(dashed lines) used for this study.\footnote{%
Nominally, there is a small gap in projected sensitivity at around $m_S\approx1$\,GeV and $\sin\theta\approx5\cdot10^{-5}$
between the future exclusion power of the HL LHCb and the upper range of validity of the ShiP limits.
This window however, will most likely be closed by {\it i)} slightly stronger (upper) limits of 
FASER2~\cite{Beacham:2019nyx} compared
to SHiP and {\it ii)} the fact that in addition to  $B^+ \to K^+ \mu^+ \mu^-$ the channel $B^+ \to K^+ \pi\pi$ will also be analysed
by LHCb. The corresponding limit is expected to be more stringent than our estimate around $m_S \sim 1$~GeV owing to 
the fact that the branching ratio of
$S$ into pions is strongly enhanced compared to the branching into muons in this mass range, see e.g.~\cite{Winkler:2018qyg}. 
When presenting our final results for the projected sensitivity of future experiments in this mass range, we will thus 
just use the {\it lower} sensitivity bound of SHiP, $\sin\theta\sim10^{-6}$.
}
For comparison, we also show existing limits from astrophysics; those will be discussed in more detail
in section~\ref{sec:astro}.

While we mainly concentrate on the case $m_{S}<  m_{\chi}$ as discussed above, we will also consider parameter regions
in which $m_{S} > 2 m_{\chi}$ 
and therefore invisible decays of the scalar naturally occur. In this case not all collider limits shown in 
figure~\ref{fig:particle_constraints} directly apply.  
To be specific, for $m_S=0.1$~GeV we will use the limits from E949 and NA62 as shown in 
figure~\ref{fig:particle_constraints} while for $m_S=1$~GeV the most stringent bound 
comes from the  BaBar measurement of $\text{BR}(B^+\to K^+ \bar{\nu} \nu)<1.6 \cdot 10^{-5}$~\cite{Lees:2013kla}.
Making use of the partial decay width $B \rightarrow K S$ (see e.g.~\cite{Schmidt-Hoberg:2013hba}), this translates into $\sin\theta \lesssim 6 \cdot 10^{-3}$ for $m_S \lesssim 4$~GeV.

\section{Constraints from cosmological and astrophysical probes}
\label{sec:cosmo}

\subsection{Cosmological evolution of the dark sector}
\label{sec:evolution}

In this section, we describe the full thermal evolution of the dark sector particles, $\chi$ and $S$, 
which can be qualitatively divided into five, partially overlapping stages.

\noindent$\boldsymbol{T>T_\text{dec}}\quad$ At high temperatures, the dark and the visible sector 
can be in chemical equilibrium due to the processes $\chi \bar{\chi} \leftrightarrow f \bar{f}$, 
$S \leftrightarrow f \bar{f}$ and $S S \leftrightarrow f \bar{f}$.
In that case both sectors also share the same temperature, through efficient scattering of the involved particles, 
so the temperature ratio 
\be
\xi\equiv T_S/T
\ee
is simply unity. For very small values of the mixing angle $\theta$, however, the total 
interaction rate $\Gamma_{\text{DS} \leftrightarrow \text{SM}}$ between the two sectors is never large 
enough  to bring them into thermal contact. Adding additional high-scale interactions to our model 
Lagrangian (\ref{eq:lagrangian-scalar}), on the other hand, would still allow to achieve chemical 
equilibrium for very large temperatures, without affecting the low-energy phenomenology.  
In this work, we will consider both of these possibilities and separately indicate the relevant parts of 
parameter space in our results.

\noindent$\boldsymbol{T<T_\text{dec}\quad}$ At some temperature $T_\text{dec}$ the dark 
sector decouples from the visible sector. To an approximation sufficient for our purpose,
this happens when  the total interaction rate equals the Hubble rate,
\begin{equation}
    \Gamma_{\text{DS} \leftrightarrow \text{SM}}(T_\text{dec}) \simeq H(T_\text{dec})\,.
    \label{eq:Tdec}
\end{equation}
This relation, in other words, allows to determine $T_\text{dec}$ as a function of our model
parameters ($m_\chi$, $m_S$, $g_\chi$ and $\theta$).
In practice we compute $\Gamma_{\text{DS} \leftrightarrow \text{SM}}$ only for 
the three number-changing reactions that keep up chemical equilibrium as mentioned in the previous 
paragraph (${T> T_\text{dec}}$); elastic scattering $Sf\leftrightarrow Sf$
will enforce kinetic equilibrium to be maintained slightly longer -- but this is a small effect given that 
both scattering partners are relativistic around $T_\text{dec}$.
After decoupling the scalar mediators still retain a thermal distribution with temperature $T_S$
as long as they are relativistic
(while non-relativistic scalars start to build up a chemical potential, even if a large quartic 
coupling $\lambda_S$ can keep them in local thermal equilibrium).
Moreover, for sufficiently large dark couplings $g_\chi$, they are also kept in thermal equilibrium
with the DM particles. Taken together, this leads to separate entropy conservation in the dark and 
visible sectors,  and hence a temperature ratio that changes with 
the respective number of effective entropy degrees of freedom  as
\be
\label{eq:xiT}
\xi (T)= 
\frac{\left[{g_*^\mathrm{SM}}(T)/{g_*^\mathrm{DS}(T)}\right]^\frac13}
 {\left[{g_*^\mathrm{SM}}(T_\mathrm{dec})/{g_*^\mathrm{DS}(T_\mathrm{dec})}\right]^\frac13}\,.
\ee
It is worth stressing that this equation only holds as long as {\it i)} the scalars are still relativistic
and {\it ii)} entropy is actually conserved, i.e.~before $S$ has decayed.

\noindent$\boldsymbol{T>T_\text{fo}}\quad$ Above a certain temperature $T_\text{fo}$, 
the DM particles will typically be in {\it chemical} equilibrium with the mediators and/or the SM heat bath.
The former is achieved via the annihilation process $\chi \bar{\chi}\leftrightarrow SS$, while the latter 
is only relevant for $T>T_\text{dec}$ and happens dominantly through the $s$-channel process 
$\chi \bar{\chi}\leftrightarrow f\bar f$. As the temperature approaches $T_\text{fo}$, the DM number density
freezes out and thereby sets the relic abundance of $\chi$ in the usual way. 
We numerically calculate 
the relic abundance with \textsc{DarkSUSY}~\cite{Bringmann:2018lay} including the Sommerfeld
enhancement of the annihilation rate, by modifying the implemented dark sector 
mediator model (\textsc{vdSIDM}) such as to fully include the temperature evolution of $\xi(T)$ as specified in eq.~(\ref{eq:xiT});
while the $t/u$-channel annihilation processes are identical to that model, we update the $s$-channel
annihilation rate to the one relevant for our case,
\be
 \sigma v_{M\o l} = \frac{g_\chi^2}{2}\frac{\sqrt{s}\,\Gamma_S(\sqrt{s}) }{(s-m_S^2)^2 + m_S^2 \Gamma_S^2} 
 \frac{{s-4m_\chi^2}}{s-2m_\chi^2}\,,
\ee
where $\Gamma_S$ represents the total width of $S$, and $\Gamma_S(\sqrt{s})$ the width to SM particles
assuming that $S$ had a mass of $\sqrt{s}$ rather than $m_S$.
Assuming that DM is entirely produced via thermal freeze-out thus essentially fixes the dark coupling $g_\chi$ as a function
of the other model parameters. For our final results we will both use this assumption and demonstrate
how the constraints on the model are affected if this assumption is relaxed (thus allowing for alternative DM production
mechanisms).\\
 If $m_S\gtrsim m_\chi$ only $s$-channel annihilation
is kinematically accessible; obtaining the correct DM abundance via thermal 
freeze-out would then require larger values of $\sin\theta$ than compatible with the constraints presented in 
section~\ref{sec:S_constraints}~\cite{Krnjaic:2015mbs}. 
For $0.1\lesssim m_S/m_\chi\lesssim1$, 
on the other hand, the freeze-out process would involve {\it two} 
species that are no longer in chemical equilibrium, and where eq.~(\ref{eq:xiT}) no longer applies.
As such a situation would require a dedicated analysis, we will in the following 
leave this small part of the parameter space unexplored.

\noindent$\boldsymbol{T < T_\text{fo}\quad}$ After the freeze-out of the dark matter particle, the mediator simply 
acts as an additional contribution to the energy density -- until it decays to SM particles. Both stages have an impact on BBN,
as discussed below. The corresponding 
lifetime of the scalar depends on the available decay channels and we adopt the results from 
ref.~\cite{Fradette:2017sdd} in the following.

Let us, finally, stress again that,
depending on the values of masses and couplings,
DM freeze-out can in principle happen both before ($T_\text{fo}>T_\text{dec}$) and after ($T_\text{fo}<T_\text{dec}$) the decoupling of the two sectors.
In this work we will restrict ourselves to light mediators 
with $m_S\leq0.1\,m_\chi$ when discussing thermal DM production (see discussion above). In this case, taking into
account the constraints on $\sin\theta$ that result from direct searches for $S$, it turns out that we
are always in the domain of $T_\text{fo}<T_\text{dec}$. Ultimately, this is a consequence of the 
Yukawa structure of the dark sector coupling, and the fact that we restrict our analysis to light DM.

\subsection{Big Bang Nucleosynthesis}
\label{sec:bbn}
In this section, we calculate BBN constraints for the scalar portal model using the formalism that was developed in \cite{Hufnagel:2018bjp, Hufnagel:2017dgo, Depta:2019lbe,Depta:2020wmr},
carefully taking into account the cosmological evolution of the dark sector as described in section \ref{sec:evolution}. 
Specifically, in order to use the model-independent constraints from ref.~\cite{Hufnagel:2018bjp}, 
we have to evaluate the values of $\tau_S$, $T_\text{fo}$ and $\xi(T_\text{fo})$ for every combination 
of $m_S, m_\chi\;\text{and}\;\sin\theta$, thereby fixing $g_\chi$ by the requirement of reproducing the observed relic abundance 
as described above.
We then confront the predicted abundances of light nuclei in each parameter point 
to the following set of observed primordial abundance ratios~\cite{Tanabashi:2018oca}:
\begin{align}
& \mathcal{Y}_\text{p} \quad & (2.45 \pm 0.03) \times 10^{-1} \label{eq:Yp_abundance} \; \,,\\ 
& \text{D}/{}^1\text{H} \quad & (2.569 \pm 0.027) \times 10^{-5} \label{eq:D_abundance} \;\,.
\end{align}
(Here $\mathcal{Y}_\text{p}$ denotes as usual the primordial mass fraction of $^4$He; we find that the requirement of obtaining the 
correct nuclear abundance ratio for ${}^3\text{He}/\text{D}$ leads to weaker limits than the above two constraints
for all parts of parameter space.)
Theoretical uncertainties associated to the nuclear rates entering our calculation are taken into account by running 
AlterBBN v1.4 \cite{Arbey:2011nf, Arbey:2018zfh} in three different modes corresponding to high, low and central 
values for the relevant rates. We then derive 95\% C.L. bounds by combining the observational and theoretical 
errors as described in more detail in Refs.~\cite{Hufnagel:2017dgo, Hufnagel:2018bjp}.

\begin{figure}[H]
    \centering
    \includegraphics[width=0.48\textwidth]{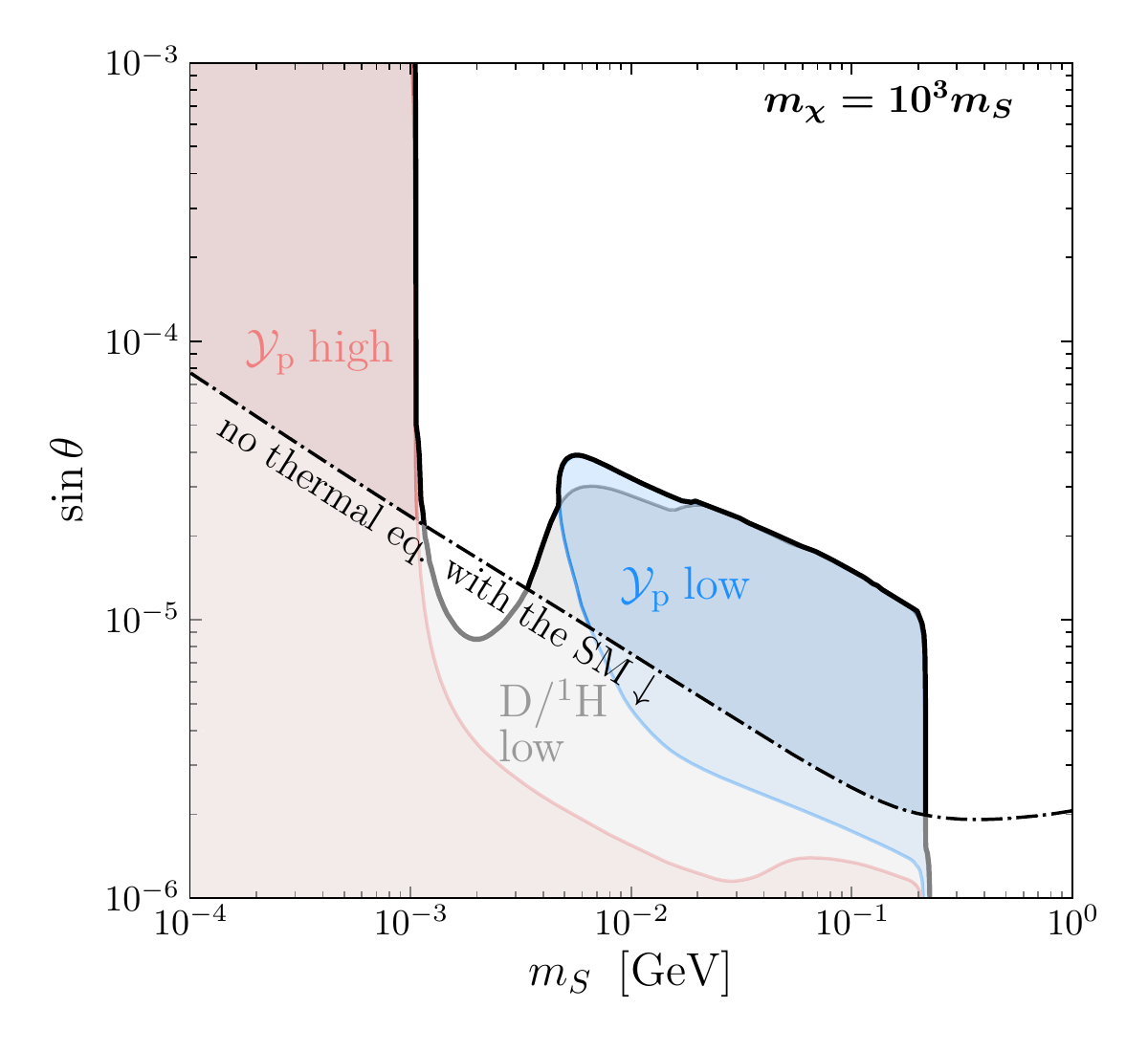}
    \includegraphics[width=0.48\textwidth]{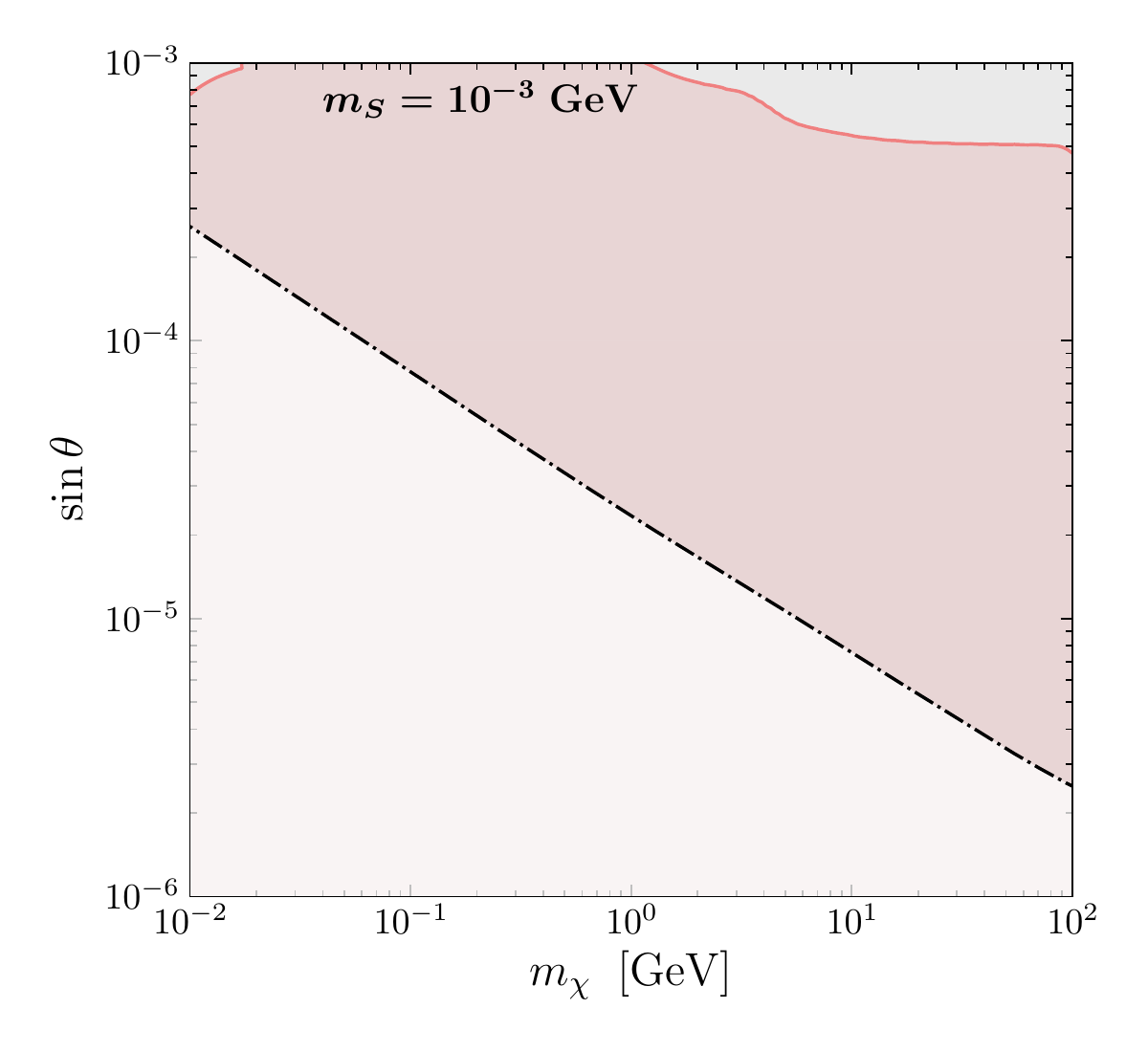}
    \includegraphics[width=0.48\textwidth]{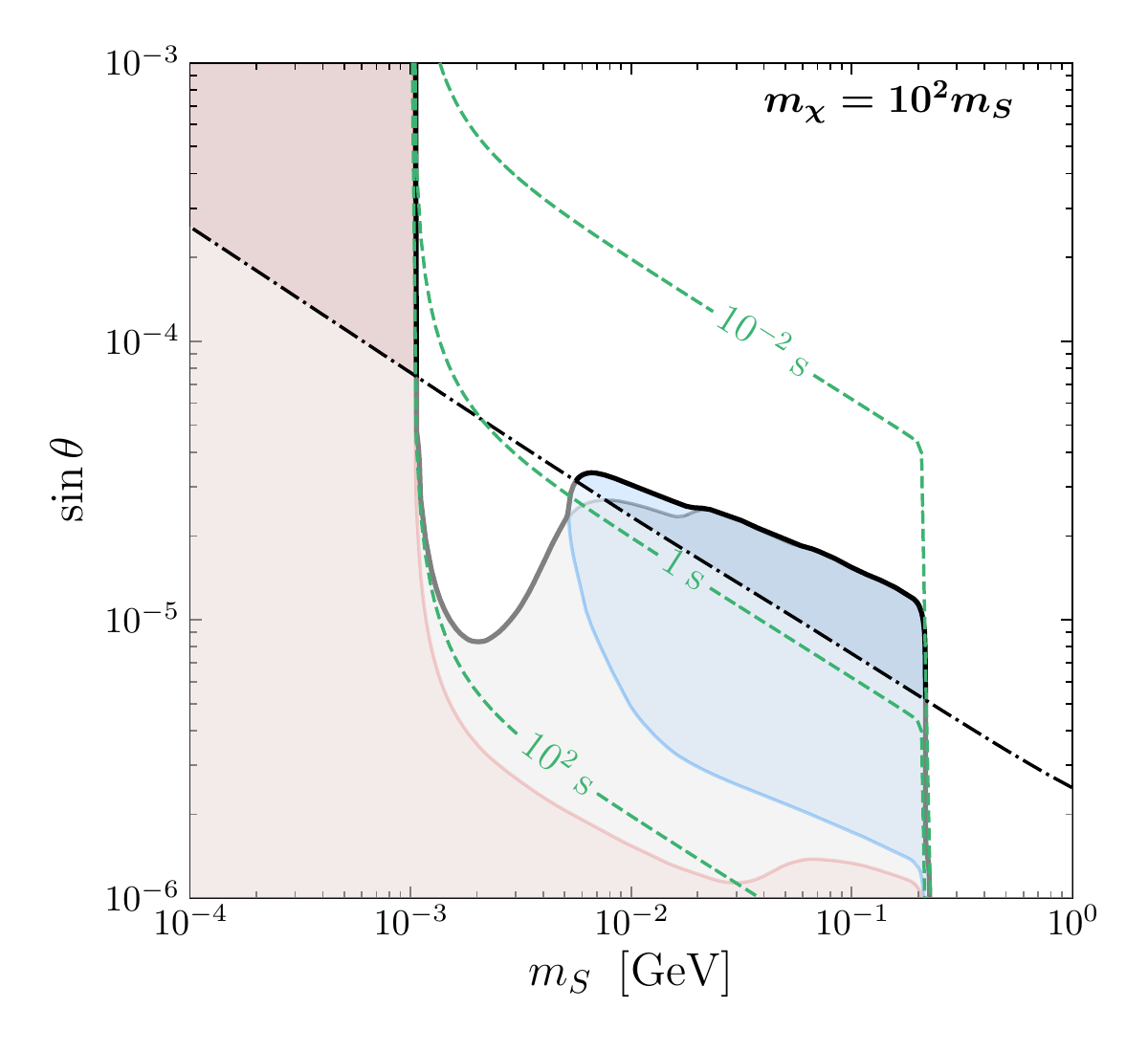}
    \includegraphics[width=0.48\textwidth]{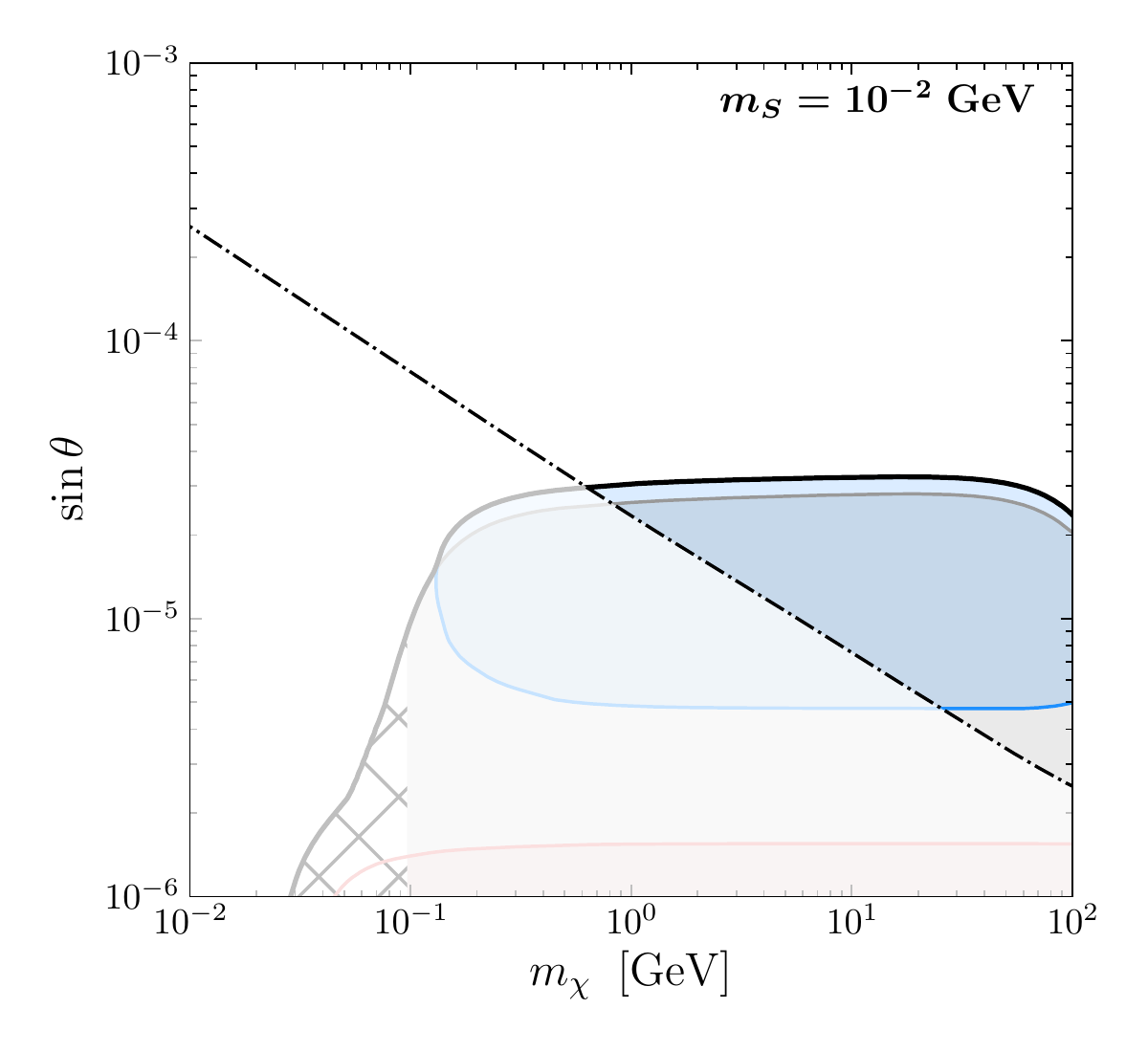}
    \includegraphics[width=0.48\textwidth]{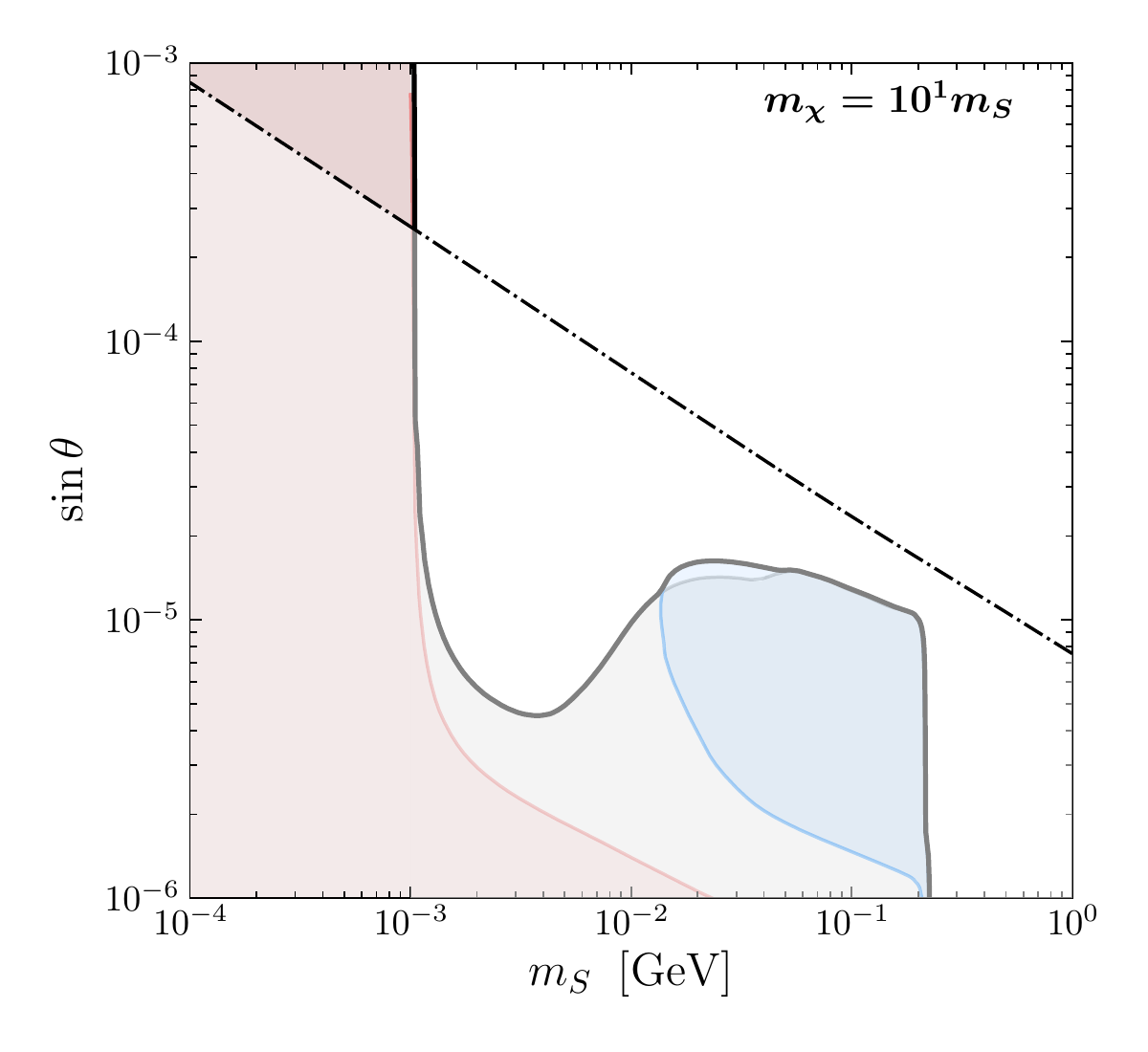}
    \includegraphics[width=0.48\textwidth]{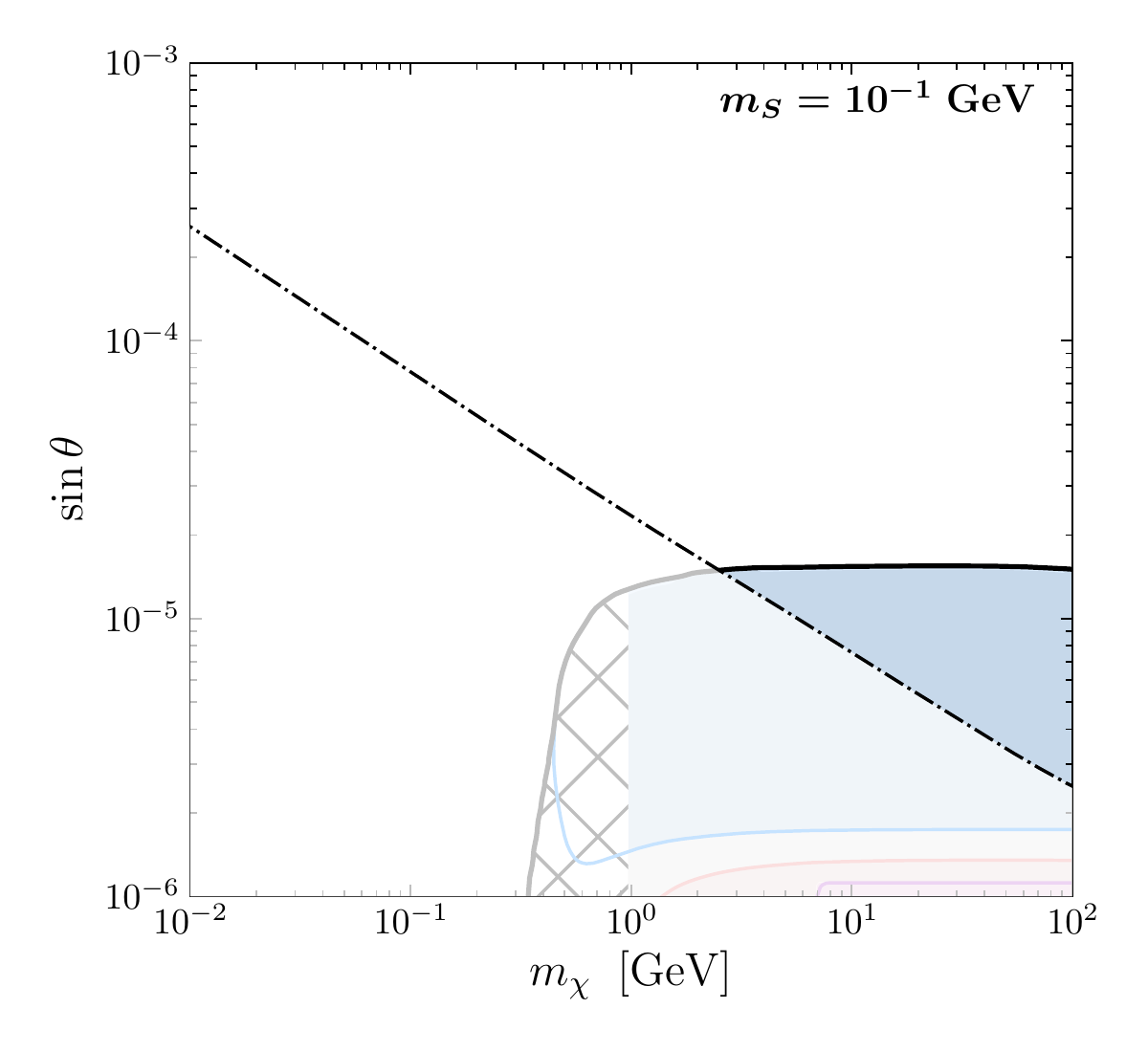}
    \caption{Limits from BBN as a function of $m_S$ and $\sin\theta$ for fixed ratios $m_S/m_\chi$ (left panels) and 
    as a function of $m_\chi$ for fixed masses $m_S$ (right panels). Below the dashed black line, the Higgs portal by 
    itself is insufficient to ever thermalise the dark sector with the SM. In addition to the overall limit (solid black line), 
    we also separately show the regions of parameter space which are excluded due to D underproduction (grey), 
    D overproduction (purple)
    and/or ${}^4\text{He}$ underproduction (blue), ${}^4\text{He}$ overproduction (pink). For 
    $m_S \gtrsim 0.1 m_\chi$ the thermal evolution is not fully captured by our calculation and thus the limits are only approximate, as indicated by the hatch pattern.}
    \label{fig:bbn_limits}
\end{figure}

In figure~\ref{fig:bbn_limits} we present the constraints from BBN as a function of $m_S$ and $\sin\theta$ for 
fixed mass ratios $m_\chi/m_S$ (panels on the left) and as a function of $m_\chi$ and $\sin\theta$ for fixed 
mediator masses $m_S$ (panels on the right). The solid black line indicates the overall limit. In addition we also 
show which nuclear abundance causes an exclusion in which part of parameter space. In the pink (blue) 
region the ${}^4\text{He}$ abundance is too large (too small) compared to the observationally inferred values, 
while the constraints due to an under- and overproduction of deuterium are shown in grey and purple, respectively. 
In addition, we show the lifetime of $S$ as a function of $\sin\theta$ for reference (green dashed lines);
the fact that we identify excluded regions with $\tau_S < 1\,$s implies that the often 
adopted `conservative BBN limit' of $\tau_S = 1\,$s is not always conservative.

It can be seen in the figure
that the limits depend significantly on the value of $m_S$ as this quantity determines the possible decay 
channels of the mediator and therefore the lifetime, while the dependence on the dark matter mass
is more indirect (but still not negligible) via its impact on the temperature ratio $\xi$.   
More concretely, we can distinguish the following regimes:
\begin{itemize}
\item For $m_S > 2m_\mu$ the limit on $\sin\theta$ is rather weak due to the small mediator lifetime above the muon threshold.
For $m_S>2 m_\pi$ also hadronic decays would become relevant for very small values of 
$\sin\theta$, see e.g.~ref.~\cite{Fradette:2017sdd}.\footnote{%
As the authors assume a different thermal history dominated by a large quartic coupling $\lambda_{hs}$, 
the bounds don't directly apply to our scenario and would need to be re-evaluated.
}
Overall we find that for values of $\sin\theta$ relevant to this study, the mediator already decays before the onset of BBN 
for $m_S > 2m_\mu$, therefore not causing any observable consequences for the range 
of direct detection cross sections we consider.
\item For  $2m_e <m_S < 2m_\mu$ the scalar can decay before, during or after BBN, depending on the value 
of $\sin\theta$. In this region of parameter space, the presence of the dark sector influences BBN via two 
different effects: {\it (i) an increase of the Hubble rate} due to the extra energy density of the dark sector and 
{\it (ii) entropy injection} into the SM heat bath due to scalar decays into electromagnetic ration. 
Both affect the synthesis of light elements as discussed in detail in ref.~\cite{Hufnagel:2018bjp} and are 
fully taken into account in our evaluation. For lifetimes $\tau_S \gtrsim 10^4$\,s photodisintegration 
also becomes relevant, but does not exclude any additional regions of parameter space (at least in case DM is produced via freeze-out).
\item For $m_S < 2m_e$, the scalar $S$ can decay only into photons, which leads to a drastically increased lifetime. 
Consequently, for comparably small values of $\sin\theta$, the mediator outlives the creation of the light elements, 
thereby acting as an additional relativistic degree of freedom, whose presence can be robustly excluded by current 
BBN constraints (even stronger constraints in the case of such late decays arise from photodisintegration of light elements~\cite{Poulin:2015opa,Hufnagel:2018bjp,Forestell:2018txr}
and the CMB~\cite{Poulin:2016anj}). For very large values of $\sin\theta$, on the other hand, $S$ again decays during 
BBN -- but since 
this case is strongly excluded by other considerations, c.f.~section~\ref{sec:particle}, we do not 
perform a detailed study of BBN limits in this regime.
\end{itemize}

As discussed in section \ref{sec:evolution}, for sufficiently small values of the mixing angle $\theta$, the two sectors 
will never thermalise via the Higgs portal;
this is indicated by the dashed black line in figure~\ref{fig:bbn_limits}.
The conservative BBN limits which do not make any additional assumptions about early universe cosmology 
therefore na\"ively end at this line.\footnote{%
The bounds may be significantly stronger taking into account an irreducible contribution from freeze-in 
production of either the DM particle or the mediator. In fact even a small abundance of mediators is constrained if the lifetime is such that
photodisintegration is relevant. A detailed exploration of this parameter region is left for future work.
Also, as stated before, we assume $\lambda_{hs}\simeq 0$. Sizeable values for this quartic inter-sector coupling
could further shift the region of thermalisation to smaller values of $\sin\theta$.
} 
While the overall limit looks very similar for a given scalar mass $m_S$, the thermalisation line is rather sensitive to 
the DM mass $m_\chi$ which directly translates into rather different conservative BBN limits for the different cases.
For the calculation of the BBN limits below this line we assume $\xi(T\rightarrow \infty)=1$, i.e.~that both sectors were 
thermalised at very large temperatures by some additional processes that are not covered by the model Lagrangian 
in eq.~(\ref{eq:lagrangian-scalar}). If the two sector never thermalised, on the other hand, the bound would depend on the initial temperature ratio $\xi$ (or ratio of energy densities).
For $\xi(T \to \infty)=0$ only the freeze-in contribution 
would remain. Nevertheless, even in this case stringent bounds from photodisintegration and the CMB are
expected for sizeable regions of parameter space. 
To indicate this additional uncertainty the BBN limits below the thermalisation line have 
a lighter shading.

\subsection{Dark matter self-interactions}
\label{sec:SIDM}

The exchange of a scalar particle generates an attractive Yukawa potential between two DM particles, resulting 
in a self-interaction rate that strongly depends on the couplings $g_\chi$, the DM and mediator masses $m_\chi$ and 
$m_S$, and the relative velocity $v$ of the scattering DM particles. For the range of parameters we are interested in here,
in particular, these interactions typically show a characteristic resonant structure, resulting in a large enhancement or suppression
of the momentum transfer cross section $\sigma_T$ when varying, e.g., the dark coupling $g_\chi$. 
In this regime, analytic expressions are available that result from restricting the analysis to 
 $s$-wave scattering and approximating  the Yukawa potential by a Hulth\'en potential~\cite{Tulin:2013teo}. 
 While these expressions result in a reasonable estimate for height and location of {\it resonances} in $\sigma_T$, we find that 
 they significantly underestimate the numerical value of $\sigma_T$ in the vicinity of {\it anti-resonances} (see also~\cite{Chu:2019awd}). 
 In our analysis, we thus always solve the underlying  Schr\"odinger equation for the full Yukawa potential numerically, 
 including also higher partial waves. We do so by following the treatment 
 in ref.~\cite{Kahlhoefer:2017umn}, thus also correctly taking into account the indistinguishability of DM particles 
 in the definition of the momentum transfer cross section.

In the cosmological concordance model, DM is successfully described as a collision-less fluid. In fact,
astrophysical observations from dwarf galaxy to cluster scales stringently limit how much the properties 
of the putative DM particles can deviate from this simple picture (for a review, see ref.~\cite{Tulin:2017ara}).
Here we adopt
\be
\label{eq:SIDMconstraint}
 \sigma_T/m_\chi < 1\,\mathrm{cm^2}/\mathrm{g}
\ee
as a fiducial maximal value for the allowed effective momentum transfer cross section
at a relative DM velocity of $v_\chi^\mathrm{rel}=1000$\,km/s. This corresponds 
roughly to the constraint inferred from the observation of colliding galaxy 
clusters~\cite{Markevitch:2003at,Randall:2007ph,Tulin:2017ara}, which are highly DM-dominated 
systems and hence often argued to be less prone to modelling uncertainties of the baryonic component.
Another advantage is that these observations more directly constrain the DM self-interaction rate, while the 
widely used translation of individual halo properties -- like their core size -- to bounds on $\sigma_T$ is subject 
to a non-negligible intrinsic modelling uncertainty (for a recent discussion, see ref.~\cite{Sokolenko:2018noz}).\footnote{%
Observations of dwarf galaxies, and their translation to limits on $\sigma_T$, are prone to much larger
uncertainties~\cite{deNaray:2009xj,Valli:2017ktb,Bondarenko:2017rfu}. 
On the other hand, for light mediators the effective self-interaction rate is 
considerably stronger in these systems than in galaxy clusters. Adopting for example
$\sigma_T/m_\chi < 100\,\mathrm{cm^2}/\mathrm{g}$ for relative DM velocities of $v_\chi^\mathrm{rel}=30$\,km/s, 
as a relatively conservative fiducial constraint, would lead to stronger constraints 
than eq.~(\ref{eq:SIDMconstraint}) only for $m_S\lesssim10^{-3} m_\chi$.
}
We note that astrophysical observables do not depend on $\sigma_T$ alone, 
but may also depend on the frequency of 
scatterings~\cite{Kahlhoefer:2013dca}. For the case of a light mediator as studied here, this can be substantially 
different from a contact-like (isotropic) DM scattering, which is usually assumed in $N$-body simulations 
(see however Refs.~\cite{Robertson:2016qef,Kummer:2017bhr,Kummer:2019yrb} for first studies including angular 
dependent and frequent scatterings). 
On the other hand, bounds on the self-interaction rate have been reported that are significantly stronger
than the fiducial maximal value(s) of $\sigma_T$ that we adopt in our 
analysis~\cite{Kaplinghat:2015aga,Elbert:2016dbb,Bondarenko:2017rfu,Harvey:2018uwf}.
Overall, taking into account the above mentioned uncertainties, 
we expect that eq.~(\ref{eq:SIDMconstraint}) 
leads to realistic constraints on the dark coupling $g_\chi$.

Due to the characteristic resonant structure of $\sigma_T$,  the inversion of these limits to 
constraints on $g_\chi$ is in general not unique.
For given values of $m_\chi$ and $m_S$, in particular, there is always a maximal value $g_\chi^\mathrm{min}$ such
that eq.~(\ref{eq:SIDMconstraint}) is satisfied for {\it all} values of $g_\chi<g_\chi^\mathrm{min}$. 
In the resonant regime, however,  it is possible to hit anti-resonances, and thus to {\it decrease} the cross section 
by {\it increasing} the coupling beyond $g_\chi^\mathrm{min}$. In other words, there can be further -- sometimes only 
very narrow -- parameter ranges with $g_\chi\in(g_\chi^\mathrm{min},g_\chi^\mathrm{max})$ that 
satisfy eq.~(\ref{eq:SIDMconstraint}). Here, $g_\chi^\mathrm{max}$
denotes the {\it maximal} value for which the self-interaction constraint can in principle be satisfied, due to the 
appearance of anti-resonances in the scattering amplitude. 
Requiring $g_\chi<g_\chi^\mathrm{max}$ is
thus the most conservative way of implementing the self-interaction constraint, but it neglects the fact that many values
of $g_\chi<g_\chi^\mathrm{max}$ are actually excluded; requiring $g_\chi<g_\chi^\mathrm{min}$
is more aggressive, but in some sense more generic (because it applies even outside the range of
couplings where anti-resonances appear).
To reflect this situation, we will in the following show results for both sets of constraints independently.
We note that numerically it is straight-forward to determine $g_\chi^\mathrm{min}$ and $g_\chi^\mathrm{max}$ 
because the anti-resonances are much less pronounced than what the analytic approximation
for  $s$-wave scattering~\cite{Tulin:2013teo} would suggest.

\subsection{Further astrophysical and cosmological bounds}
\label{sec:astro}

Weakly coupled light particles can be copiously produced in the interior of stars or in the hot core of a supernova (SN) via their
interactions with electrons or nucleons. For sufficiently weak couplings these particles escape the celestial body without
further interactions and therefore constitute a new energy loss mechanism which is constrained by observations.
We take the resulting limits on $\sin\theta$ from red giants (RG) and horizontal branch stars (HB) from ref.~\cite{Hardy:2016kme}.
The bound from SN 1987A which extends to larger masses because of the correspondingly larger core temperature 
we take from ref.~\cite{Winkler:2018qyg}, noting that this is an order of magnitude estimate with inherently large uncertainties. 
The lower boundary of this limit is determined by estimating the additional energy loss due to escaping scalars which would 
shorten the observed neutrino pulse.  For $m_S<2m_\chi$ the SN limit does not extend to arbitrarily large couplings due to 
efficient trapping of light scalars inside the SN; for larger scalar masses, on the other hand, there is no upper boundary 
of the limits because the scalar will decay invisibly to DM particles that escape the SN without interacting. 
These constraints are shown as grey areas in the $\sin\theta - m_S$ plane in figure~\ref{fig:particle_constraints}, 
where we have used a hatched filling style for the SN bounds (assuming $m_S<2m_\chi$) to stress their intrinsic uncertainty.

We remind that the usually very strong CMB bounds on annihilating light DM can be evaded in this model because
the annihilation proceeds via a $p$-wave and is hence strongly velocity-suppressed (also, there are no remaining light degrees of
freedom that would change the expansion rate at these times). While the elastic scattering of DM with SM particles is 
enhanced for light mediators $S$, the coupling of $S$ to photons is still not sufficient to prolong kinetic decoupling
until times where an appreciable cutoff in the matter power spectrum, and hence a potential imprint in e.g.~Ly-$\alpha$ data, 
would be expected (see ref.~\cite{Bringmann:2016ilk} for a more detailed discussion).

\section{Results}
\label{sec:results}

As motivated in the introduction, we now want to combine the various constraints discussed
in the previous sections in the `direct detection plane', i.e.~as bounds on $\sigma_{\rm SI}$ 
as a function of the DM mass. For a given value of the scalar mass $m_S$ (and fixed $m_\chi$)
only the invisible Higgs decay constrains the same combination of parameters ($\sin\theta\cdot g_\chi$) 
that enters the expression for $\sigma_{\rm SI}$, c.f.~eq.~(\ref{eq:sigsifull}). In all other cases,
we thus need to combine two types of observations to constrain these parameters individually.
As the particle physics constraints discussed in section~\ref{sec:particle}, but also the astrophysical constraints
from section \ref{sec:astro}, are essentially insensitive to $g_\chi$, a handle on the dark coupling has to be provided
by cosmology. Concretely, we will distinguish three versions of cosmological constraints (roughly ordered by 
decreasingly strong underlying assumptions):

\begin{itemize}
\item {\bf Cosmo 1} {\it (`thermal production')}. The present dark matter abundance 
can be fully explained by the production of $\chi$ particles via freeze-out in the early universe,
as detailed in section \ref{sec:evolution},
which requires dark and visible sector to have been in thermal equilibrium at some point.
No further interactions than those specified in eq.~(\ref{eq:lagrangian-scalar}) are assumed.\footnote{%
For $m_S\gtrsim m_\chi$ the relic density actually depends on the same combination of couplings 
($\sin\theta\cdot g_\chi$) as  direct detection rates; as noted before, it is not possible to obtain the 
correct relic density and at the same time satisfy existing bounds on $\theta$ in this case. 
}
\item {\bf Cosmo 2} {\it (`generic self-interactions')}. No connection between the dark coupling $g_\chi$ and the DM abundance is assumed, allowing for other DM production mechanisms. `Generic' 
constraints on DM self-interactions are adopted, as detailed in section \ref{sec:SIDM}, i.e.~we assume that 
$g_\chi$ does not lie close to an anti-resonance in the elastic scattering cross section.
\item {\bf Cosmo 3} {\it (`conservative self-interactions')}. As Cosmo 2, but implementing the most conservative 
constraints from DM self-interactions; larger values of $g_\chi$ would thus violate eq.~(\ref{eq:SIDMconstraint}) 
even when finely tuned to lie on an anti-resonance.
\end{itemize}
BBN constraints are tightly coupled to the assumed thermal history, so for those we will always assume thermally
produced DM (`Cosmo 1'). Finally, we always adopt a hard `perturbative unitarity limit' of $g_\chi< \sqrt{4\pi}$ in case the respective cosmological constraint would be weaker.

\begin{figure}[H]
\begin{center}
\mbox{}\\[-0.7cm]
\hspace*{-0.3cm}
\includegraphics[width=0.45\textwidth]{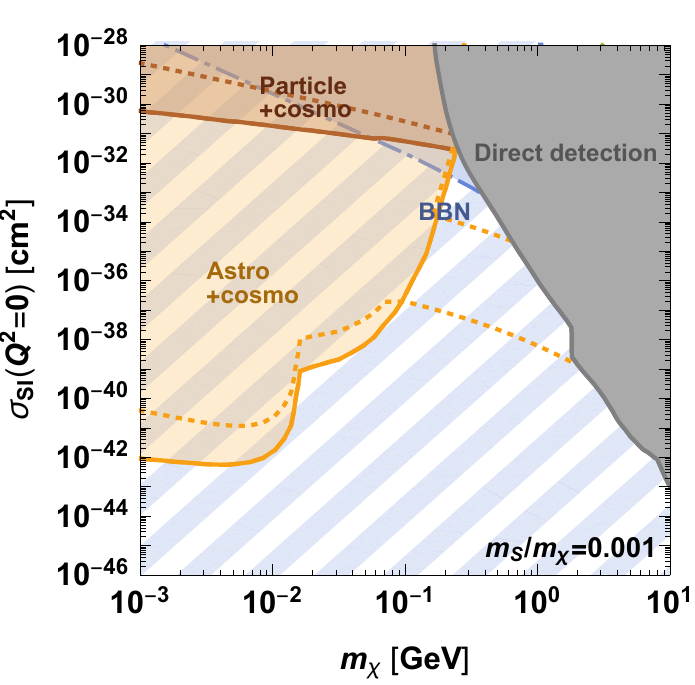}
\hspace*{0.5cm}
\includegraphics[width=0.45\textwidth]{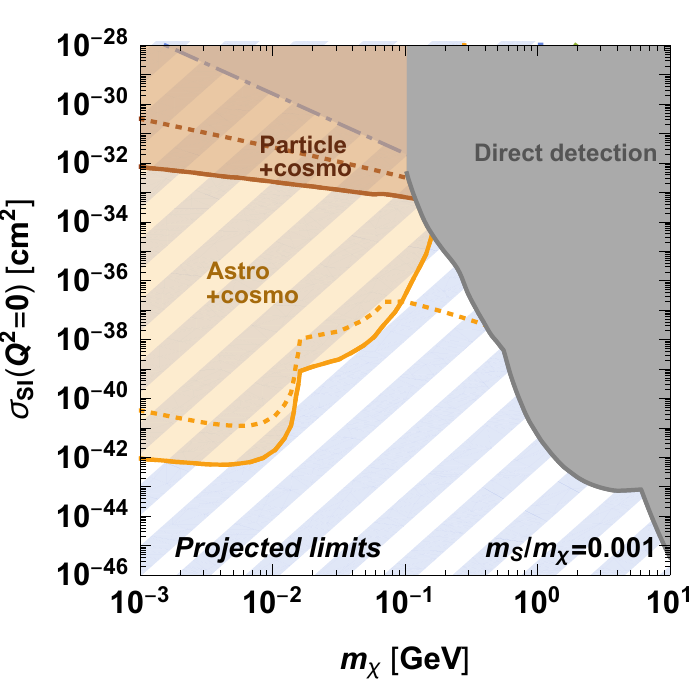}\\
\hspace*{-0.3cm}
\includegraphics[width=0.45\textwidth]{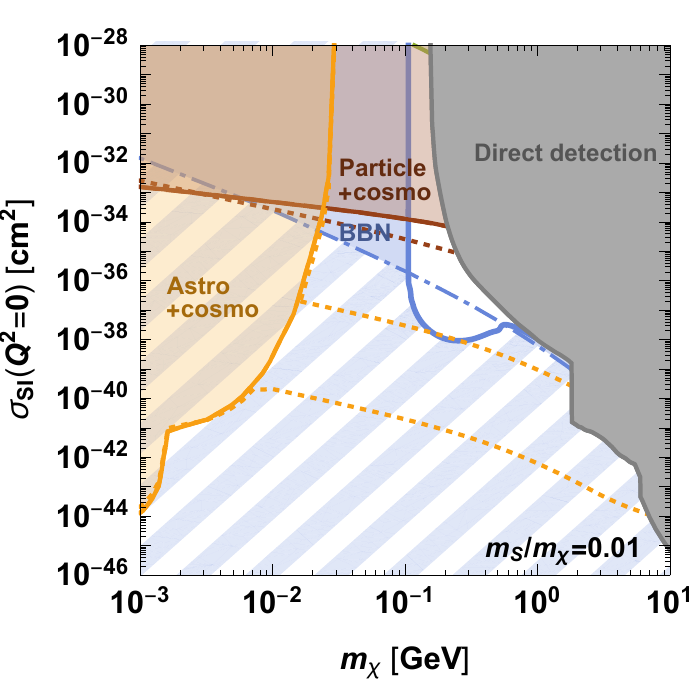}
\hspace*{0.5cm}
\includegraphics[width=0.45\textwidth]{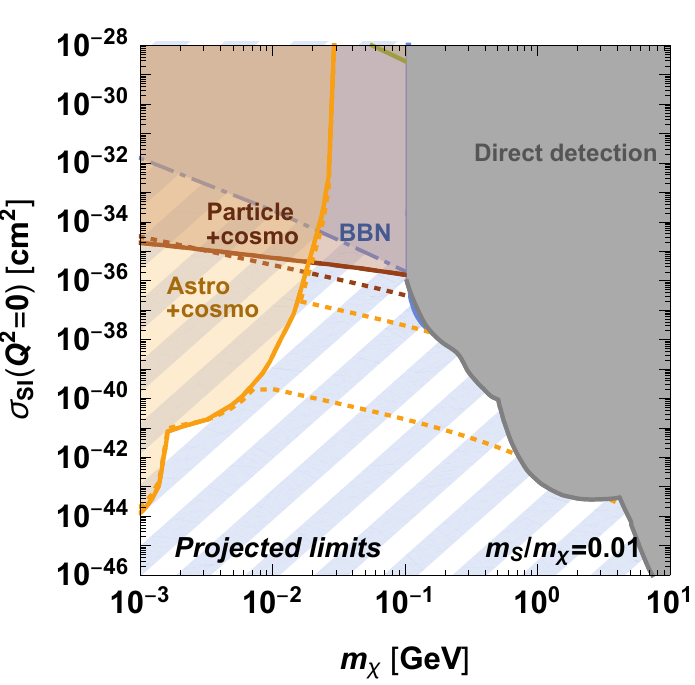}\\
\hspace*{-0.3cm}
\includegraphics[width=0.45\textwidth]{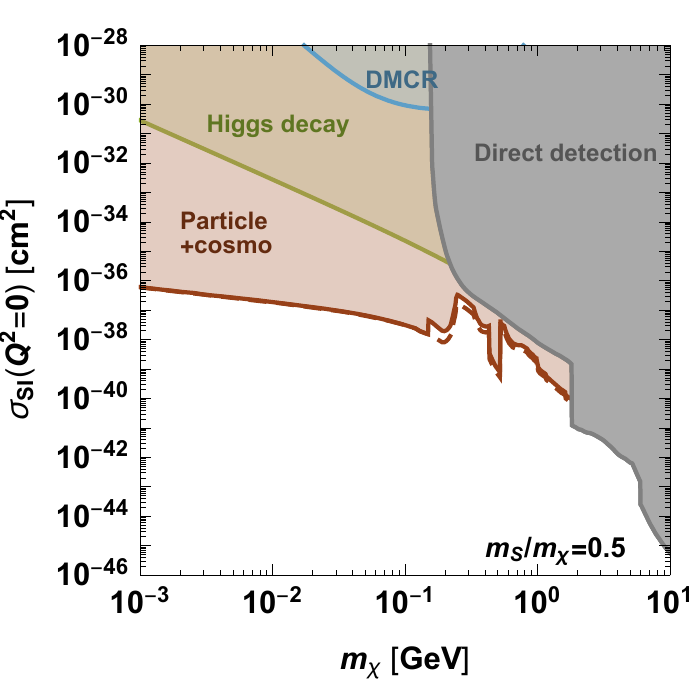}
\hspace*{0.5cm}
\includegraphics[width=0.45\textwidth]{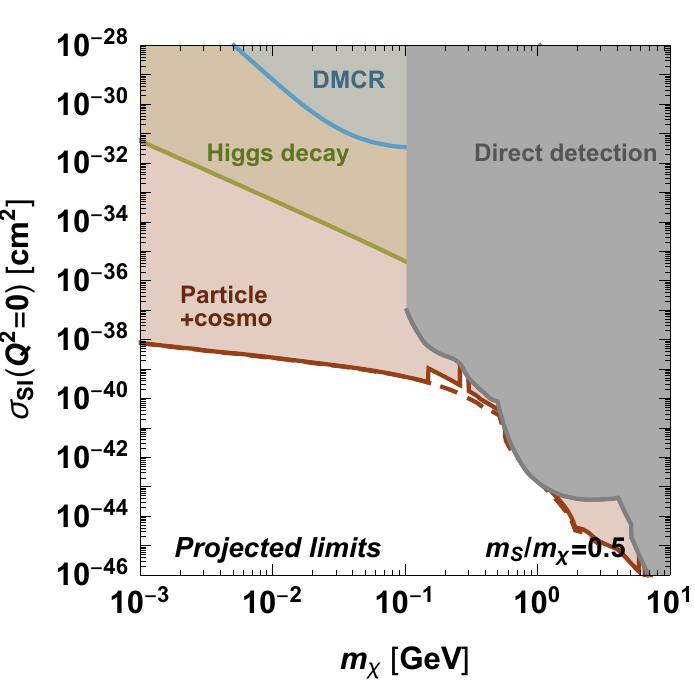}\\[-1cm]
\end{center}
\caption{Current (left) and projected (right) limits on the elastic scattering cross section with nucleons in the zero-momentum
transfer limit, for fixed scalar to DM mass ratios $m_S/m_\chi$ that do not allow invisible decays of $S$. 
For astrophysical and particle
physics limits combined with cosmological limits, dotted lines assume thermal DM
 production via freeze-out (`Cosmo 1'), dashed lines instead implement generic DM self-interaction
 constraints (`Cosmo 2') while solid lines result from tuning $g_\chi$ such as to resonantly
 suppress the DM self-scattering rate (`Cosmo 3'). See text for further details. 
 }
\label{fig:results_ratio}
\end{figure}

In figure~\ref{fig:results_ratio} we show our results for selected mass ratios $m_S/m_\chi<2$ where invisible decays
of the scalar are kinematically forbidden.
In each case, the left column displays current limits while the right column displays
projected limits. Conventional direct detection limits (rescaled from figure~\ref{Fig:DDexp}) are 
shown as grey areas. Current limits from cosmic-ray upscattering (figure~\ref{fig:CRDM_constraints}) 
are shown in light blue; for the projected limits we take the sensitivity of DARWIN~\cite{Aalbers:2016jon},
based on the assumption that the recoil threshold can be lowered to 1\,keV.
Limits from invisible Higgs decay, c.f.~eqs.~(\ref{eq:higgslimit}) and (\ref{eq:higgsfuture}), 
are shown in green. In red, we combine the particle physics limits shown in figure~\ref{fig:particle_constraints}, 
while the yellow
lines show a combination of the astrophysical limits discussed in section \ref{sec:astro}.\footnote{%
From the shape of these limits, the potentially controversial part that derives from SN 
bounds is clearly discernible. We note that only in the case of `Cosmo 1', which {\it fixes} $g_\chi$ by the 
requirement of obtaining the correct relic density, the range in $\theta$ excluded in figure~\ref{fig:particle_constraints}
(for a given value of $m_S$) translates to a correspondingly excluded range of $\sigma_\mathrm{SI}$. If instead there
is only an upper limit on $g_\chi$, as in the case of `Cosmo 2' and `Cosmo 3', the direct detection cross section
$\sigma_\mathrm{SI}\propto g_\chi^2\sin^2\theta$ remains essentially unconstrained by this bound 
(in other words, while $g_\chi$ still cannot be chosen so small that $\sin\theta>1$, for a given value of $\sigma_\mathrm{SI}$,
this only results in a limit too weak to be visible in the figure). 
}
The various ways of implementing cosmological limits are indicated with dotted
lines (`Cosmo 1'), dashed lines (`Cosmo 2') and solid lines (`Cosmo 3'), respectively.
Given the difficulties in accurately computing the thermal evolution of the dark sector for 
$m_S \gtrsim 0.1\,m_\chi$, see the discussion in section \ref{sec:evolution}, we do not display 
`Cosmo 1' limits in this regime. For the case of BBN limits (shaded blue areas), we also indicate 
(as in figure~\ref{fig:bbn_limits})
the parameter region where additional high-scale interactions would be required to thermalise
the dark and visible sectors in the very early universe; BBN limits that rest on this additional
assumption are plotted with a hatched filling style. (Note that, compared to figure~\ref{fig:bbn_limits}, BBN
limits appear to have a stronger dependence on $m_\chi$ here; this is exclusively because 
the relic density constraint fixes $g_\chi$ as a function of  $m_\chi$.)

The figure nicely illustrates the complementarity of the different
approaches to test models with light mediators. If DM is thermally produced, in particular, current 
bounds already reduce the remaining parameter space for sub-GeV DM to a relatively small region of 
mediator masses above a few MeV, and mass ratios $0.01\lesssim m_S/m_\chi\lesssim 0.1$ (see also~\cite{Krnjaic:2019dzc}
for a discussion of BBN limits in a similar context). Here
it is worth commenting that BBN limits far below the thermalisation line essentially just constrain 
the {\it assumed} high-scale temperature ratio between the two sectors;
in this sense they simply exclude this additional assumption and are completely independent of the specific model Lagrangian 
stated in eq.~(\ref{eq:lagrangian-scalar}).
On the other hand the robust bounds may significantly extend below the thermalisation line taking into account 
the irreducible contribution from freeze-in production. 

Even if no assumptions about the thermal history and production of
DM is made, on the other hand, the combination of the limits displayed in figure~\ref{fig:particle_constraints} 
with those stemming
from DM self-interactions translate to highly competitive constraints on $\sigma_\mathrm{SI}$. 
For mediator masses close to the DM mass, in particular, those constraints already fully cover the 
expected reach of upcoming direct detection experiments. Interestingly, we arrive at this conclusion 
independently of which set of SIDM constraints we implement (`Cosmo 2' or `Cosmo 3'); let us stress, 
however, that the limits presented in figure~\ref{fig:results_ratio} indeed strongly depend on fully solving
the Schr\"odinger equation to obtain the self-interaction cross section $\sigma_T$ in the resonant regime,
rather than following standard practice and simply adopting analytic results for $s$-wave scattering.
From the perspective of future direct detection experiments, the most interesting parameter 
range to be probed -- fully orthogonal to what can be tested by particle physics experiments --
is the sub-GeV DM range combined with scalar masses significantly lighter than DM (but heavier than 
about 0.2\,MeV, where astrophysical limits start to dominate).

\begin{figure}[H]
\begin{center}
\mbox{}\\[-0.7cm]
\hspace*{-0.3cm}
\includegraphics[width=0.45\textwidth]{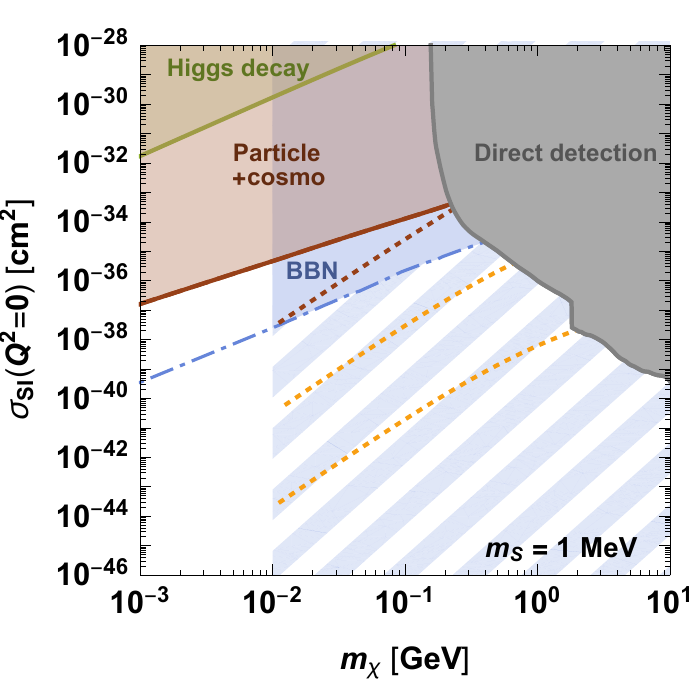}
\hspace*{0.5cm}
\includegraphics[width=0.45\textwidth]{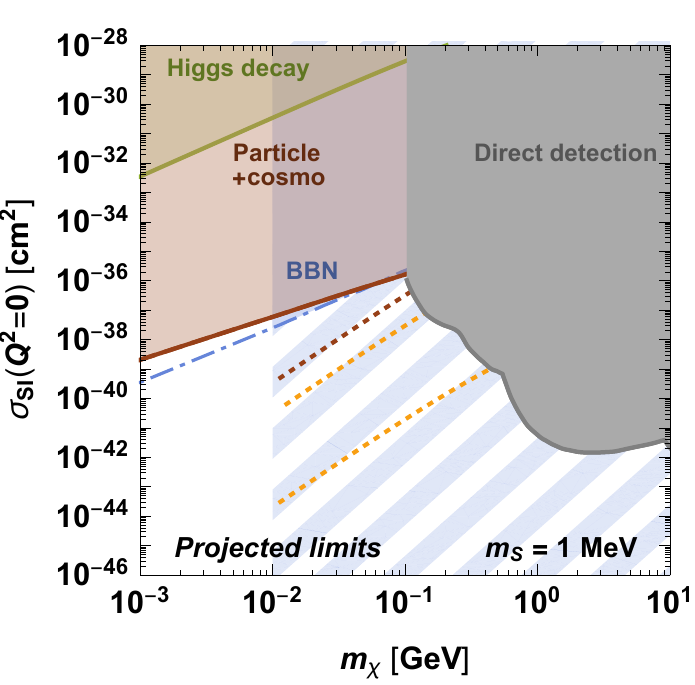}\\
\hspace*{-0.3cm}
\includegraphics[width=0.45\textwidth]{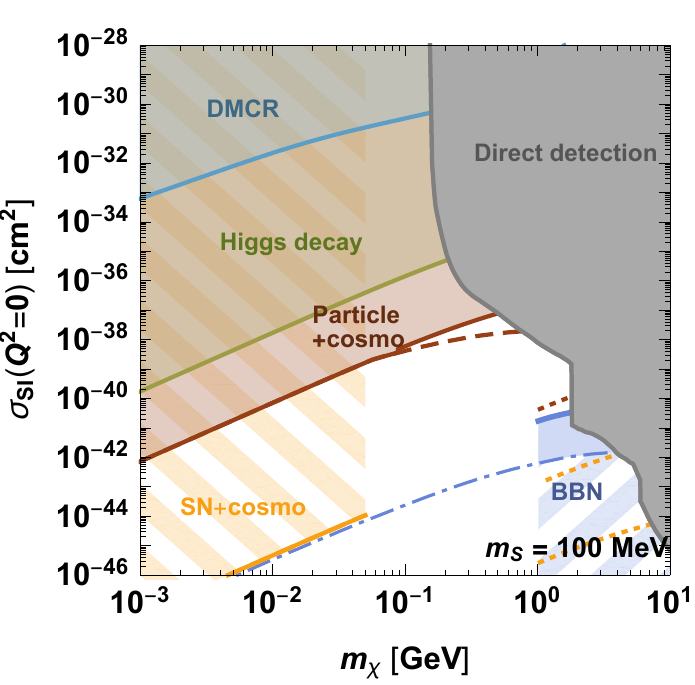}
\hspace*{0.5cm}
\includegraphics[width=0.45\textwidth]{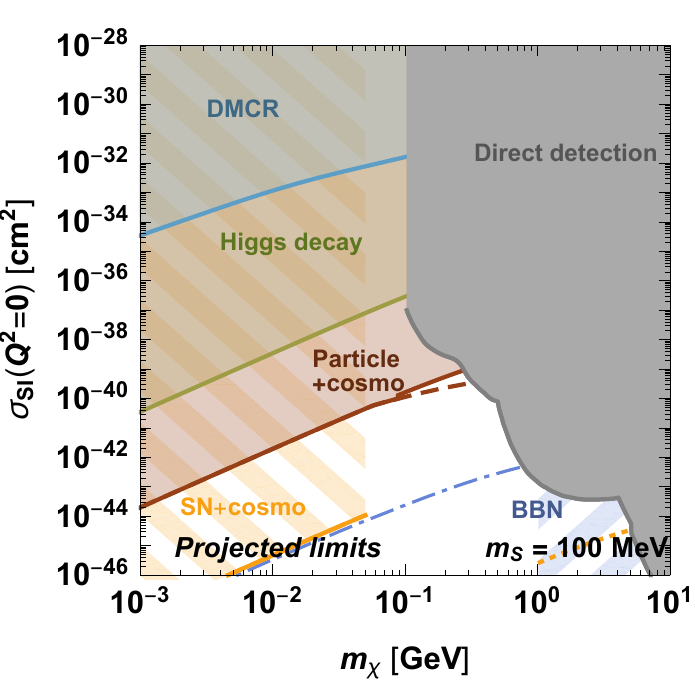}\\
\hspace*{-0.3cm}
\includegraphics[width=0.45\textwidth]{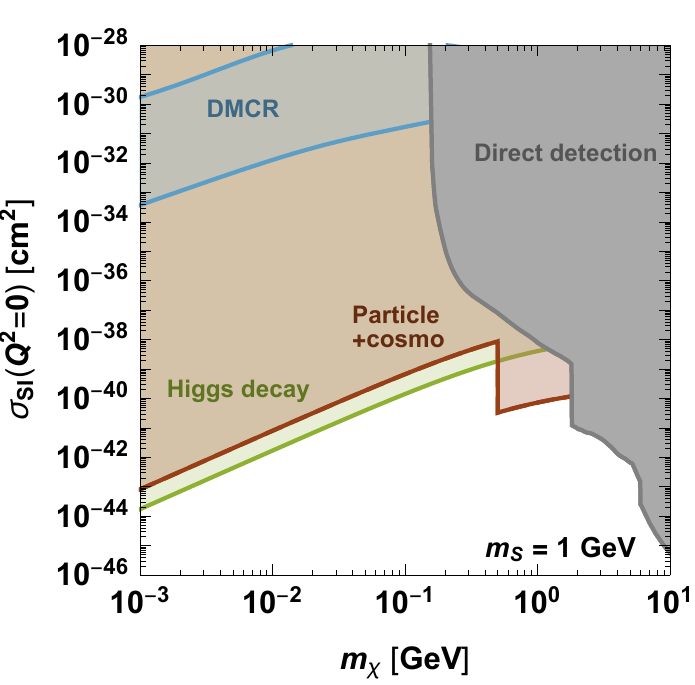}
\hspace*{0.5cm}
\includegraphics[width=0.45\textwidth]{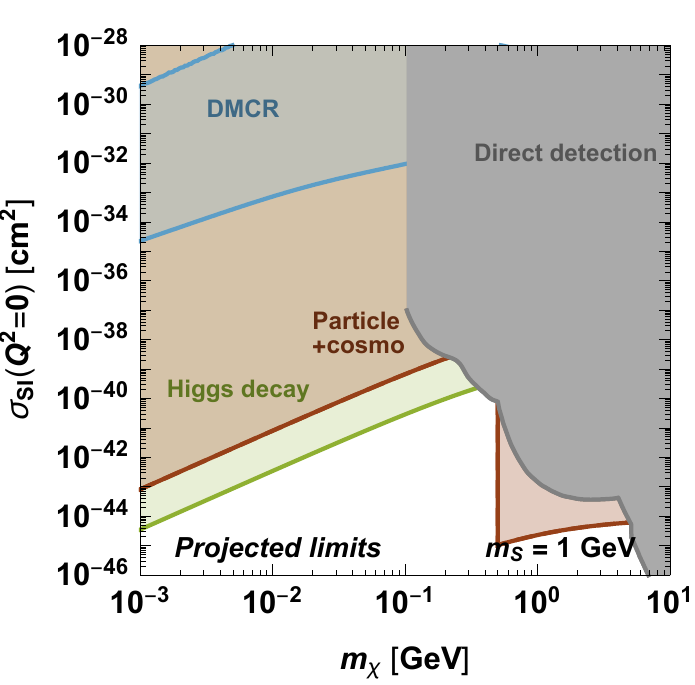}\\[-1cm]
\end{center}
\caption{Same as figure~\ref{fig:results_ratio}, for fixed scalar masses $m_S$. 
As in that figure, we do not display limits related to thermal production of DM
in the kinematic regime where $m_S \gtrsim 0.1\,m_\chi$.
}
\label{fig:results_mass}
\end{figure}

In order to add a slightly different angle to the above discussion, we show in figure~\ref{fig:results_mass} 
the same constraints for selected fixed scalar masses $m_S$ instead.
This includes kinematical situations with $m_S>2m_\chi$ where the scalar can decay very efficiently to two
DM particles, i.e.~through an invisible channel. As discussed in section~\ref{sec:S_constraints}, the particle 
physics constraints hence need to be adapted correspondingly, and we thus only keep those 
limits shown in figure~\ref{fig:particle_constraints} that are still relevant in this situation 
(and add that from BaBar~\cite{Lees:2013kla} for the 
case of $m_S=1$~GeV).\footnote{This transition between visible and invisible decays of $S$ is the reason for the 
apparent `jump' in the `particle+cosmo' limits for the case of $m_S=1$~GeV.} For invisible decays, furthermore, there is also no upper boundary to the area 
excluded by energy loss arguments in supernovae (as in figure~\ref{fig:particle_constraints}). 
This implies that for $2m_\chi<m_S\lesssim 0.2$\,GeV, unlike the situation in figure~\ref{fig:results_ratio} for 
visible decays, the combination of SN bounds and SIDM
constraints indeed does combine to a very competitive bound on $\sigma_\mathrm{SI}$ (though,
as discussed in section \ref{sec:astro}, SN limits should be interpreted with care).


\newpage
While the limits from cosmic-ray upscattered DM now become more visible, it is clear that they are never
competitive to other limits in this model. Also limits from invisible Higgs decay, while significantly more
stringent, are rarely strong enough to be competitive; this would change only with a dedicated Higgs factory 
like the ILC. In general, one can say that astrophysical, particle physics and direct detection limits
probe the parameter space from rather orthogonal directions. While astrophysical constraints
are most relevant for small DM (or, rather, mediator) masses, direct detection experiments place the strongest
limits for large DM masses. The $m_\chi$-dependence of constraints on $\sigma_\mathrm{SI}$ stemming 
from particle physics, on the other hand, is somewhat weaker. Consequently, particle physics (combined
with cosmological input) tends to place the most relevant constraints on the model at intermediate DM masses
(for the sub-GeV range that we consider here), 
and the most promising avenue for direct DM searches appears to lie in lowering the detection threshold,
even slightly, in a way that compromises the overall sensitivity as little as possible
(this, in other words, will test more of the so far unprobed parameter space than focussing on very low thresholds
at the expense of overall sensitivity).


\section{Discussion and conclusions}
\label{sec:conc}

In this work we have considered the prospects of future direct detection experiments to test uncharted 
parameter space for light (sub-GeV) dark matter. 
It is natural in this context to expect additional light particles mediating the interactions between
dark matter and the target nuclei in order to achieve a sufficiently large scattering cross section. 
To alleviate the strong cosmological bounds from the CMB we have concentrated on a scenario in 
which dark matter couples via a scalar mediator (with coupling $g_\chi$) such that dark matter annihilations proceed 
via $p$-wave and are therefore strongly suppressed at the time of the CMB.
This allows the dark matter relic abundance to be set via thermal freeze-out, although other production mechanisms are possible,
and our bounds also apply to more general cases. 
We assume that couplings to Standard Model states are induced by
the well-known Higgs portal with mixing angle $\theta$.

The DM scattering cross section off nuclei is then proportional to the product of couplings, $g_\chi^2 \cdot \sin^2\theta$.
To map out the available parameter space we evaluate and compile the relevant limits both on $\sin\theta$ and on $g_\chi$
from current and near future particle physics experiments, BBN, astrophysics and cosmological considerations.
We also show limits on light DM particles upscattered by cosmic rays, which turn out never to be competitive in the model
considered here. In our analysis we paid special attention to cosmological constraints which, while requiring certain 
assumptions, cannot be avoided altogether in a given model. Indeed, they provide quite in general the missing link 
to translate a  variety of constraints on portal models to constraints on the scattering cross section relevant for 
direct dark matter searches. 

In our analysis we update and carefully extend previous recent work similar in 
spirit~\cite{Krnjaic:2015mbs,Kainulainen:2015sva,Knapen:2017xzo,Evans:2017kti}. 
Concretely, we re-scale direct detection limits according to the full $Q^2$ dependence in section~\ref{sec:dd}, 
and present a genuinely new treatment of cosmic-ray up-scattered DM for such a case -- which, as we stress, 
is applicable also to other scenarios that invoke $Q^2$-dependent scattering. 
We present updated invisible Higgs decay constraints on this specific model 
(in section~\ref{sec:higgsinv}), and add genuinely new estimates of the future LHCb and NA62 sensitivities 
(see appendix) to our compilation of up-to-date particle physics constraints and projected sensitivities 
in section~\ref{sec:S_constraints}.  A significant refinement compared to what is typically done in the 
literature is further that we consistently implement the cosmological evolution of this scenario in full detail 
(section~\ref{sec:evolution}), which we use both for precision calculations of the relic density and 
a careful evaluation of BBN constraints (going well beyond the standard procedure of simply translating
model-independent constraints to the model at hand). 
We also point out that the self-interaction cross section (section~\ref{sec:SIDM}) has to take into 
account the (in)distinguishability of external particles and needs to be evaluated beyond the $s$-wave 
approximation to avoid the appearance of artificially deep anti-resonances and hence overly weak constraints; 
correcting for this, we instead find that DM self-interactions
generically lead to limits comparable to those resulting from the \emph{assumption} of DM thermally produced
via the freeze-out mechanism.

Overall we find that, almost independently of the dark matter production mechanism, strong bounds on the
maximally possible nuclear scattering rate exist for large regions of parameter space. Nevertheless, some 
regions remain safe from the combination of existing (or expected) constraints from accelerators, astrophysics and cosmology,
motivating the development and construction of future direct detection experiments which could explore these regions.
This not only requires low thresholds for the recoil energies, but at the same time sensitivities better than what is presently
achievable at dark matter masses around 1\,GeV.

\bigskip
\acknowledgments

We would like to thank Christopher Cappiello, Frederik Depta, Babette D\"obrich, Camilo Garcia-Cely, Timon Emken, 
Maxim Pospelov, Giuseppe Ruggiero and Martin Winkler for helpful discussions.
This work is supported by the ERC Starting Grant `NewAve' (638528), the ERC Advanced Grant `NuBSM' (694896), the 
Deutsche Forschungsgemeinschaft under Germany's Excellence Strategy -- EXC 2121 `Quantum Universe' --
390833306 as well as by the Netherlands Science Foundation (NWO/OCW). We further thank the Erwin Schr\"odinger International Institute for hospitality while this work was completed.


\newpage
\appendix

\section{Estimate of future LHCb sensitivity}
\label{app:lhcb}

\begin{figure}[t]
\begin{center}
\includegraphics[width=0.6\textwidth]{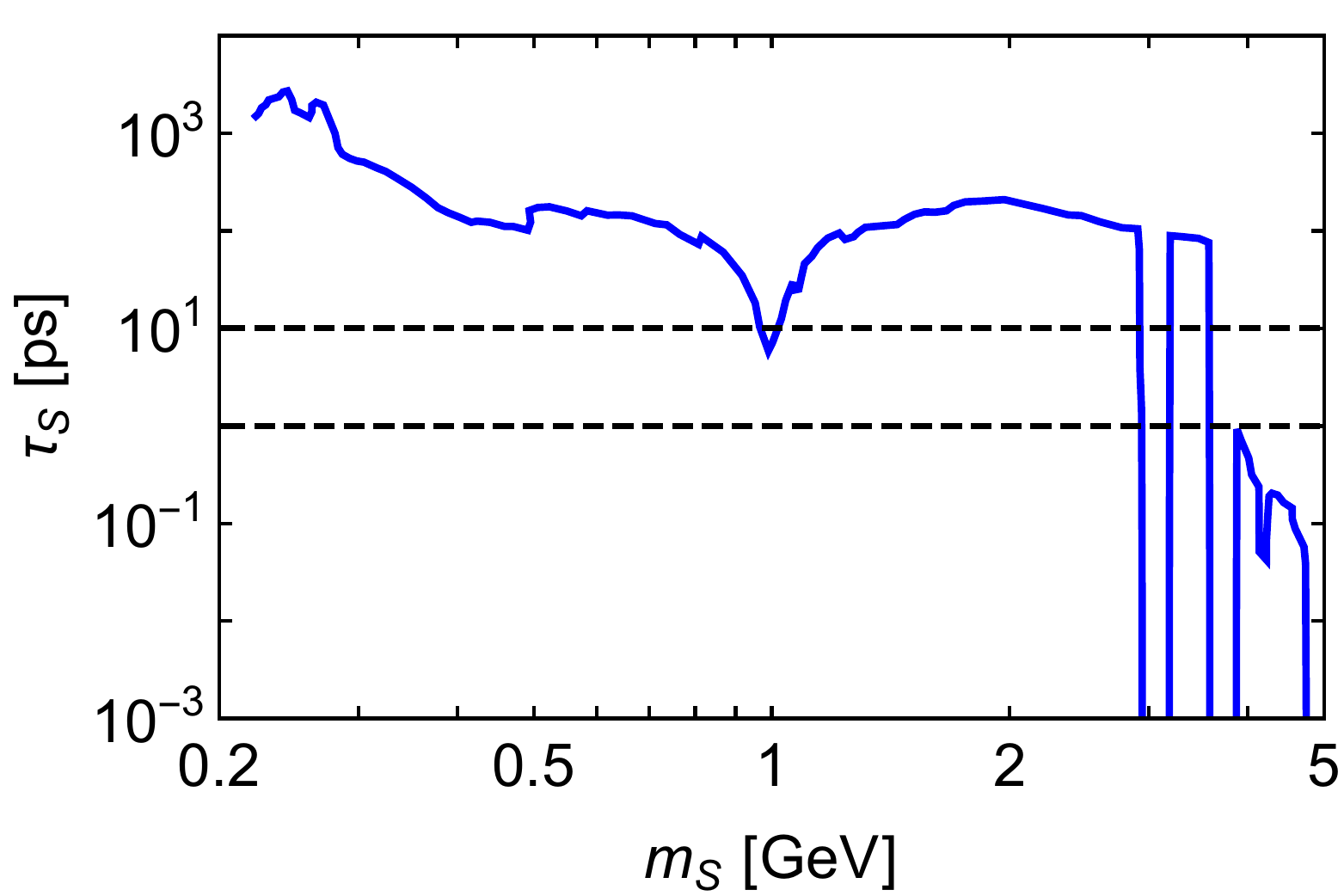}
\caption{Lifetime of the scalar particle $S$ as a function of its mass, thereby fixing $\sin\theta$ to the lower bound of the current LHCb sensitivity as shown in figure~\ref{fig:astro-accel} (blue line). 
The black dashed lines correspond to lifetimes of $1$~ps and $10$~ps, thus indicating the borders between the prompt region ($\tau_S < 1\;\mathrm{ps}$), the intermediate region ($1\;\mathrm{ps} < \tau_S < 10\;\mathrm{ps}$) and the large displacement region ($\tau_S > 10\;\mathrm{ps}$). See text for further details.}
\label{fig:LHCb-tauS}
\end{center}
\end{figure}

The loop-induced rare decay $B^+ \to K^+ S \to  K^+ \mu^+ \mu^-$ with muons in the final state can potentially be observed at LHCb. 
For the values of $\theta$ currently probed, the lifetime of $S$ can become significant on detector scales, such that a search for displaced vertices
will significantly enhance the sensitivity~\cite{Schmidt-Hoberg:2013hba}.
Such an analysis has been performed by ref.~\cite{Aaij:2016qsm} 
for a dataset with collision energy $\sqrt{s} = 7$ and $8$~TeV and integrated luminosity $\mathcal{L}_0 = 3\text{ fb}^{-1}$.
For this analysis, the parameter space of the scalar was divided into {\it (i)} a prompt region with the lifetime of the scalar $\tau_S < 1$~ps, {\it (ii)} an intermediate region with $1\text{ pc}<\tau_S<10$~pc and 
{\it (iii)} a large displacement region for $\tau_S>10$~pc. Background events were expected in the first two regions, while the last region was background free.
In figure~\ref{fig:LHCb-tauS} we show $\tau_S$, fixing $\sin\theta$ to the lower bound of the current sensitivity of the LHCb experiment. 
We conclude that no background is expected for $m_S<3.7$~GeV (which is the region of interest to us), while for higher masses we need to consider a non-zero background contribution.

To estimate the sensitivity of a similar analysis in the high-luminosity (HL) phase of the LHC, we assume the total integrated luminosity of LHCb to be $\mathcal{L}_{\text{HL}} = 300\text{ fb}^{-1}$ and 
the centre-of-mass energy to be $\sqrt{s} = 13$~TeV. The corresponding increase of the number of produced $B$ mesons in the direction of LHCb can be estimated as
\begin{equation}
    \mathcal{R} = \frac{\mathcal{L}_{\text{HL}} \cdot \sigma_{13}(pp\to B^+ +X)}{\mathcal{L}_0 \cdot \sigma_{8}(pp\to B^+ +X)} \approx 162.2\,,
\end{equation}
where $\sigma_{13/8}(pp\to B^+ +X)$ is the production cross section of $B^+$ mesons which fly into direction of the LHCb detector for energies $13$ and $8$~TeV respectively. 
We estimated these cross sections using FONNL (Fixed Order + Next-to-Leading Logarithms) -- a model for calculating the single inclusive heavy quark production cross section, 
see~\cite{Cacciari:1998it,Cacciari:2001td,Cacciari:2012ny,Cacciari:2015fta} for details.

For the region in which background events are expected, we assume for simplicity that the number of background events also increases by the factor $\mathcal{R}$. 
We estimate the future sensitivity as
\begin{equation}
    \theta^2_{\text{future}}(m_S) = \frac{\theta^2_{\text{current}}(m_S)}{\sqrt{\mathcal{R}}}\,.
\end{equation}

For the case of large displacements, while no background events are expected, we need to take into account the probability of the scalar to decay inside the region where displaced vertices can be observed, 
$l_{\min} \le l_{\text{decay}} \le l_{\max}$ with $l_{\min}=3$~mm and  $l_{\max} = 0.6$~m, \cite{Aaij:2016qsm}. This probability can be written as
\begin{equation}
    P_{\text{decay}}(\theta) = e^{-l_{\min}/l_{\text{decay}}(\theta)} - e^{-l_{\max}/l_{\text{decay}}(\theta)}\,,
\end{equation}
where $l_{\text{decay}} = c \gamma_S \tau_S$ is decay length of the scalar in the lab frame and $\gamma_S$ is the corresponding Lorentz factor. 
We estimate the average Lorentz factor of the scalar to $S$ be (see appendix C in~\cite{Bondarenko:2019yob})
\begin{equation}
    \gamma_S = \gamma_{S,\text{rest}} \frac{E_B}{m_B}\,,
\end{equation}
where $\gamma_{S,\text{rest}}$ is the Lorentz factor of $S$ in the rest frame of the $B$-meson. 
The average energy of the $B$-mesons in the direction of LHCb we take from FONNL, $E_B \approx 80$~GeV for both centre-of-mass energies. 
Taking everything together, we estimate the future sensitivity in this regime as
\begin{equation}
    \theta^2_{\text{future}}(m_S) P_{\text{decay}}(\theta_{\text{future}}(m_S))= \frac{1}{\mathcal{R}}
    \theta^2_{\text{current}}(m_S)
    P_{\text{decay}}(\theta_{\text{current}}(m_S))\,.
\end{equation}
which is shown in figure~\ref{fig:particle_constraints}. 
For earlier estimates of the LHCb sensitivity see, e.g., Ref.~\cite{Clarke:2013aya}.

\section{Estimate of future NA62 sensitivity}
\label{app:NA62}

In this appendix we briefly describe how we estimate the sensitivity of the NA62 experiment with respect to light scalars produced in $K^{+}\to \pi^{+} S$.%
\footnote{For a sensitivity estimate of NA62 to light scalars with different coupling structure see e.g.~\cite{Krnjaic:2019rsv}.}
One of the main physics goals of NA62 is the measurement of the rare decay $K^{+}\to \pi^{+}\nu \bar{\nu}$, allowing for a direct determination of the $V_{td}$ CKM matrix element~\cite{NA62:2017rwk}. The observed final state is a $\pi^+$ plus missing momentum. If the scalar $S$ is sufficiently long-lived to decay outside of the
detector it would contribute to the same final state and can therefore be constrained with this search mode. 
A crucial difference between the expected signal from $K^{+}\to \pi^{+} S$ compared to the SM process $K^{+}\to \pi^{+}\nu \bar{\nu}$ is the distribution of the 
`invisible mass', which in the case of decays into $S$ peaks at the mass $m_S$ while for the SM process (as well as other SM backgrounds which contribute to this final state) the distribution is rather flat\footnote{Due to large backgrounds from $K^{+}\to \pi^+ \pi^0$ the mass range $130~\text{ MeV}<m_S<150~\text{ MeV}$ should be excluded from the analysis~\cite{NA62:2668548}.}, see e.g.~\cite{CortinaGil:2018fkc}. The number of kaons expected during LHC Run 3~\cite{Beacham:2019nyx} is estimated to be $N_K \simeq 10^{13}$, which (scaling up the results from~\cite{CortinaGil:2018fkc}) corresponds to about 35 events in the signal region with a rather flat distribution in the missing mass.

To compare this to the expected signal from $K^{+}\to \pi^{+} S$ we have to take into account the corresponding branching ratio as well as the total selection efficiency $\epsilon$ for this process, which in general will depend on $m_S$. The expected number of events is then
\begin{equation}
    N_S^\text{obs} = N_K \cdot \text{BR}(K^{+}\to \pi^{+} S) \cdot \epsilon \,.
\end{equation}
For our analysis we approximate the total efficiency as $\epsilon =0.3$~\cite{na62_private}. The relevant branching ratio is given by (see e.g.~\cite{Boiarska:2019jym})
\begin{equation}
    \text{BR}(K^{+}\to \pi^{+} S) \simeq 1.85\cdot 10^{-3} \sin^2\theta \sqrt{\left( 1 - \frac{(m_S+m_{\pi})^2}{m_K^2}\right)\left( 1 - \frac{(m_S-m_{\pi})^2}{m_K^2}\right)} \,.
\end{equation}
Taking into account that the experimental resolution of the missing mass is about 1/35 of the signal region, we expect about 1 background event from SM processes after all cuts. The 95\% CL upper limit on the scalar would then correspond to $\sim 5$ events, which is what has been required for our result shown in figure~\ref{fig:particle_constraints}. 


\bibliography{CRDM}
\bibliographystyle{JHEP}

\end{document}